\documentclass[9pt,twocolumn,twoside]{osajnl}
\journal{jocn} 
\usepackage{longtable}
\usepackage{pgfplotstable}

\usepackage{subfigure}
\usepackage{adjustbox}
\usepackage{array}
\usepackage{lscape}
\DeclareUnicodeCharacter{2060}{\nolinebreak}
\usepackage{tikz}
\usepackage{soul}
\usepackage{amssymb}\usepackage{pifont}\newcommand{\cmark}{\ding{51}}\newcommand{\xmark}{\ding{55}}\usepgfplotslibrary{groupplots}
\usepackage{tikz-3dplot}
\pgfplotsset{compat=1.3}
\usepgfplotslibrary{colorbrewer}
\usetikzlibrary{plotmarks}
\pgfplotsset{colormap/Dark2,
cycle list/Dark2,
    }

\setboolean{shortarticle}{false}

\title{Topology Bench: Systematic Graph Based Benchmarking for Core Optical Networks}

\author[1,*,**]{Robin Matzner}
\author[2,*]{Akanksha Ahuja}
\author[1]{Rasoul Sadeghi}
\author[1]{Michael Doherty}
\author[1]{Alejandra Beghelli}
\author[2]{Seb J. Savory}
\author[1]{Polina Bayvel}

\affil[1]{Optical Networks Group, Department of Electronic \& Electrical Engineering, UCL (University College London), London WC1E 7JE, UK}
\affil[2]{Department of Electrical Engineering, University of Cambridge, Cambridge, CB3 0FA, United Kingdom}

\affil[*]{These authors contributed equally to this work.}
\affil[**]{Corresponding author: robin.matzner.19@ucl.ac.uk}

\begin{abstract}
Topology Bench is a comprehensive topology dataset designed to accelerate benchmarking studies in optical networks. 
The dataset, focusing on core optical networks, comprises publicly accessible and ready-to-use topologies, including (a) 105 georeferenced real-world optical networks and (b) 270,900 validated synthetic topologies.
Prior research on real-world core optical networks has been characterised by fragmented open data sources and disparate individual studies. 
Moreover, previous efforts have notably failed to provide synthetic data at a scale comparable to our present study.
Topology Bench addresses this limitation, offering a unified resource and represents a 61.5\% increase in spatially-referenced real world optical networks.
To benchmark and identify the fundamental nature of optical network topologies through the lens of graph-theoretical analysis, we analyse both real and synthetic networks using structural, spatial and spectral metrics.
Our comparative analysis identifies constraints in real optical network diversity and illustrates how synthetic networks can complement and expand the range of topologies available for use.  
Currently, topologies are selected based on subjective criteria, such as preference, data availability, or perceived suitability, leading to potential biases and limited representativeness. 
Our framework enhances the generalisability of optical network research by providing a more objective and systematic approach to topology selection.
A statistical and correlation analysis reveals the quantitative range of all of these graph metrics and the relationships between them.
Finally, we apply unsupervised machine learning to cluster real-world topologies into distinctive groups using nine optimal graph metrics using K-means. It employs a two-step optimisation process: optimal features are selected by maximising feature uniqueness through Principal Component Analysis, and the optimal number of clusters is determined by maximising decision boundary distances via Support Vector Machines.
We conclude the analysis by providing guidance on how to use such clusters to select a diverse set of topologies for future studies.
Topology Bench, openly available via \href{https://zenodo.org/records/13921775}{Zenodo} and \href{https://github.com/TopologyBench}{GitHub}, promotes accessibility, consistency, and reproducibility.
\end{abstract}

\setboolean{displaycopyright}{false} 

\begin{document}

\maketitle

\section{Introduction}
Optical networks carry the majority of digital data throughout the world, revolutionising the way we live our lives in the last 40+ years. With year-on-year exponentially increasing internet traffic, new ways of increasing the throughput of these networks is a constant challenge \cite{networking_cisco_2020}.

A long-standing research goal has been to design wavelength routed optical networks (WRON). The problem of designing the structural (how the nodes are connected) and the spatial properties of the network (length of edges) is referred to as the physical topology design (PTD) problem. These networks are implemented via wavelength division multiplexing (WDM), which uses individual wavelengths to route data over different lightpaths on the same fibre \cite{hill_wavelength_1988,chlamtac_lightpath_1992}. In WRONs, the challenge is to map a given traffic demand over specific paths and wavelengths, often referred to as the routing and wavelength assignment (RWA) problem \cite{chlamtac_lightpath_1992}. Both these optimisation problems determine the final performance of a network in terms of maximum achievable throughput, resilience and latency.

Due to evolving technology (e.g. flexible grid or multiband/core/mode networks) and varying traffic requirements (static/dynamic) in optical networks, there are a diverse set of resource allocation problems. For example, RWA on fixed-grid WRONS, routing and spectrum allocation (RSA) on flex-grid or elastic optical networks (EON) \cite{eon_gerstel_2012}, RWA/RSA on multiband \cite{Sambo2020} or multicore/mode \cite{Boffi2022, Klinkowski2018} fibre networks, network virtualisation-type problems like virtual optical network embedding (VONE) \cite{zhu_vone_2013, doherty_deep_2023}, network function virtualisation \cite{vnfc_zhu_2024}, and distributed sub-tree scheduling for multicast \cite{multicast_2024}. We refer to all of these problems as resource allocation problems from this point on.

The PTD problem and resource allocation problems are NP-hard combinatorial optimisation problems \cite{tornatore_wdm_2007, liu_physical_2008}, with numerous solutions proposed, for example, \cite{baroni_analysis_1999, chatterjee_review_2013,luo_exploring_2023, saito_cost-effective_2024}. Additionally, there has been a recent surge in machine learning (ML) applications to both the PTD problem and the resource allocation problem literature \cite{tanaka_reinforcement-learning-based_2024, cicco_deep_2022}. Due to the lack of standardised benchmarking practices, it remains difficult for researchers to compare published solutions to these problems. For instance, DeepRMSA, a highly cited ML solution for routing, modulation and spectrum assignment (RMSA), was trained and evaluated on NSFNET and COST239 topologies with link lengths scaled down by a factor of two \cite{deeprmsa}. Some authors \cite{deeprmsa_gcn_rnn, heuristic_reward_design} have continued to use the examples from DeepRMSA, while others have used the full-scale topology versions \cite{nevin_techniques_2022, nallaperuma_interpreting_2023}, and others have opted for different real-world topologies \cite{mask_rsa, quang_magc-rsa_2022}.

Synthetic topologies have been another substitute for finding topologies to test algorithms on \cite{cicco_deep_2022}. There are a handful of generative graph models for the use of optical core network research. The Gabriel Graphs, Waxman and SNR-BA models are amongst the most investigated \cite{cetinkaya_comparative_2014, pavan_generating_2010, matzner_making_2021}. However, a comprehensive published dataset of synthetically generated networks is still not available. As a result, authors must generate topologies every time they wish to work with synthetic data.  

Finally, another fundamental problem has been that published datasets have not been analysed using graph signal processing. Without this analysis, it is difficult to justify the selection of networks used for benchmarking algorithmic solutions. Topology diversity in terms of structural, spatial properties and network scale is key to demonstrate the applicability of resource allocation solutions.

In summary, there is a need within the optical networking research community to provide: (i) an open-source exhaustive real-world optical network dataset (ii) a synthetically generated dataset to complement the gaps in real networks (iii) a rigourous graph theoretical analysis of the provided topologies to allow for a diverse selection of networks to test upon.

With these requirements in mind, these are the key contributions of this paper:
\begin{enumerate}

    \item \textbf{Topology Bench Dataset:} This dataset comprises the most extensive collection of 105 real optical core networks to-date, with two subsets of 900 and 270,900 synthetic graphs, available at \href{https://zenodo.org/records/13921775}{Zenodo}. The real dataset expands options for optical network researchers by 61.5\%, providing significantly more topologies for study. The synthetic networks compliment the real networks and contain networks over different node scales, connectivities and distance scales marking the first public release of a large-scale synthetic dataset for diverse research needs.

    \item \textbf{Topology Analysis:} 
    To our knowledge, this work presents the first systematic graph-theoretical analysis of a large-scale, open-source dataset combining real and synthetic optical networks.
    Both types of networks in Topology Bench are analysed to provide researchers with a comprehensive overview of their structural, spectral and spatial properties.
    Furthermore, this work provides critical insights into the composition of Topology Bench, and the potential impact of network properties on the performance of resource allocation solutions

    \item \textbf{Topology Clustering and Selection: } 
In this study, we introduce the first categorisation of 105 real optical network topologies using unsupervised machine learning, based on their structural, spatial, and spectral properties, a novel approach that has not been explored in prior research. 
These well-defined clusters, offer researchers a powerful and objective tool for selecting topologies with diverse characteristics and follow robust benchmarking practices.

\end{enumerate}

The rest of the paper is organised as follows. Section 2 reviews previous work on network benchmarking in ML and optical networking, and identifies limitations of existing optical network datasets. Section 3 reviews the real and synthetic topologies that are part of Topology Bench and the network analysis metrics. Section 4 and 5 investigate the structural and spatial properties of real and synthetic networks respectively. Finally, section 6 investigates how to select real and synthetic topologies from these datasets.

\begin{figure*}[h!]
    \centering
    \includegraphics[width=\linewidth]{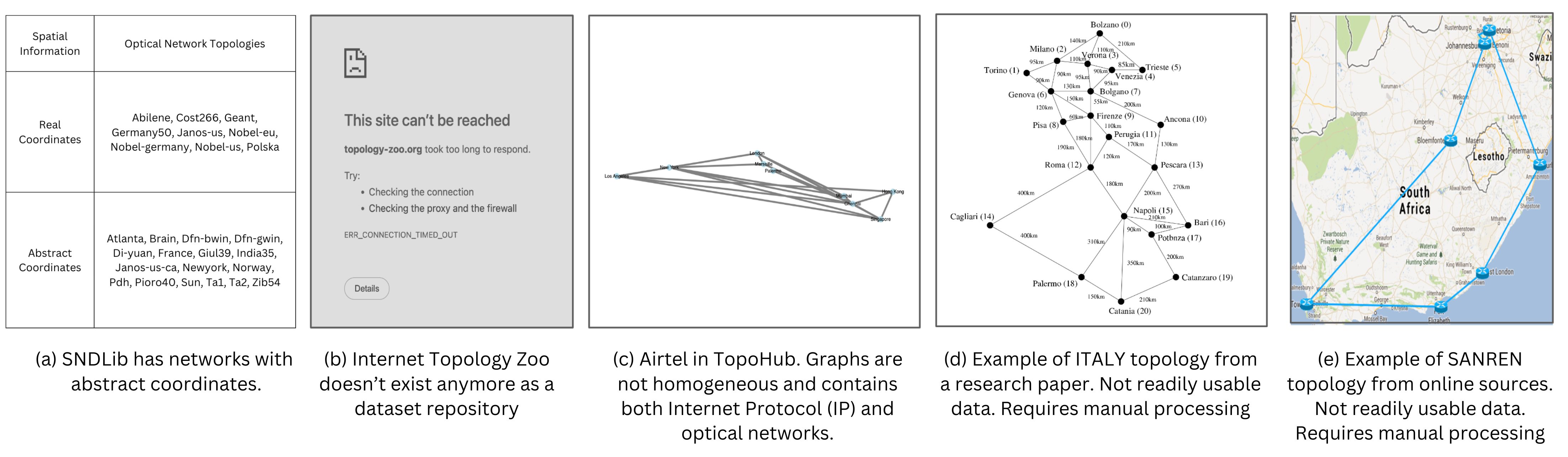}
    \caption{
    Limitations in optical network topology datasets impede accurate modelling, benchmarking, and scalable solutions.
}
    \label{fig:limit_data}
\end{figure*}

\section{Related Work}
\label{sec:related_work}
To place our work in context and motivate Topology Bench as a resource for optical networks research, this section reviews related efforts in benchmarking and dataset standardisation.

\subsection{Existing Benchmarks in Machine Learning}
Machine learning (ML) advancements in the past two decades have been significantly influenced by the availability of large-scale, open-access datasets. In computer vision, datasets such as ImageNet \cite{russakovsky_imagenet_2015} and Common Objects in Context (COCO) \cite{fleet_computer_2014} have facilitated model comparison and improvement, standardised problem formulation and evaluation metrics, enabled transfer learning, enhanced reproducibility, set industry standards, and improved fairness in ML systems. Similarly, graph-based datasets like Open Graph Benchmark (OGB) \cite{hu_open_2021} with 16 datasets each having millions of nodes and links and TUDataset \cite{morris_tudataset_2020} with 125 datasets having hundreds of nodes and links are the reason behind the rapid growth of graph machine learning. 

However, these existing graph datasets do not accurately represent the structural characteristics of real-world optical networks. Topology Bench addresses this gap by providing a comprehensive domain specific dataset specifically designed for optical network research and enables the use of novel graph representation learning techniques \cite{ahuja2024topology}.
\subsection{Existing Datasets in Optical Network Topologies}
While previous efforts have attempted to provide optical network topology datasets, they have fallen short in various aspects. SNDLib \cite{orlowski_sndlib_2010} offers 26 topologies, but over half lack real spatial location data (Figure \ref{fig:limit_data}a). The Internet Topology Zoo \cite{knight_internet_2011}, a once-valuable crowd-sourced resource, is no longer accessible as of April 18, 2024 (Figure \ref{fig:limit_data}b). Although we've archived their files, spatial information is limited to a subset of nodes and topologies. Additionally,  many topologies represent both IP and optical networks, an example of which is Airtel network (Figure \ref{fig:limit_data}c).
Recent aggregation efforts, such as TopoHub \cite{jurkiewicz_topohub_2023}, merely compile data from SNDLib and Internet Topology Zoo without addressing their inherent limitations. Other topologies found in literature are often presented only as visualisations (Figures \ref{fig:limit_data}d and \ref{fig:limit_data}e), lacking the raw data in readily usable formats necessary for reproducibility.
Critically, no single repository has previously offered a comprehensive collection of topology data that includes both geographic coordinates of nodes and link lengths—information vital for physical layer impairment-aware algorithms. Topology Bench addresses this gap by providing not only this essential data for real networks but also a set of realistic synthetic topologies.

\subsection{Existing Methods for Generating Network Topologies}
\label{sec:existing_synthetic_work}
Although real network datasets give a reference point for testing algorithms, they do not offer a wide variety of network sizes (number of nodes), network densities and distance scales. Generative graph models and synthetic datasets of optical networks provide this flexibility to investigate how topological features affect performance metrics, as well as giving algorithmic results statistical significance. In addition, they can be used to train data-intensive ML models \cite{matzner_intelligent_2023}. 

The oldest and simplest generative graph model is that of Erdös and Renyi (ER) \cite{erdos_evolution_1960}. It is a random model, that models the probability of an edge existing as a Bernoulli distribution with probability $p_{ER}$. Many simulation studies in optical networking have used this or a very similar model \cite{wu_interference-and-security-aware_2016, ashraf_spatially_2018, depizzol_feature_2018, fenger_statistical_2000, chatelain_topological_2009, baroni_wavelength_1997, higashimori_impact_2022}. However, the ER model generates graphs inherently different from those of real optical core networks in terms of nodal degree distribution and link lengths \cite{matzner_making_2021}.

The Barabasi and Albert model (BA) tackled the problem of nodal degree distribution, by using the principle of preferential attachment \cite{barabasi_emergence_1999}. This makes nodes added into the graph more likely to connect to nodes with a high degree. This model results in degree distributions that resemble a power law, resulting in many nodes with smaller degrees, and a few highly connected hub nodes. However, this model neglects the spatial properties of optical networks.

Multiple authors investigated the suitability of geometric generative graph models to generate graphs that take the spatial properties of the networks into account. The most investigated are the Gabriel Graph (GG) and Waxman model \cite{gabriel_new_1969, waxman_routing_1988}. The GG model adds an edge between two nodes if there is no other node that lies within the radius from the halfway point between these two nodes. Although the GG model was shown to give graph properties closest to the real networks in \cite{cetinkaya_comparative_2014}, the model is deterministic, meaning that for a given set of nodes there is a single Gabriel graph. This lack of flexibility makes it difficult to generate a variety of topologies. The Waxman model exponentially decays the probability of nodes to be connected according to the distance. This exponential decay according to distance is not comparable to the approximate inverse linear scaling of Signal-to-Noise Ratio in optical networks with number of spans \cite{matzner_making_2021}. Although determined to be the best model in \cite{pavan_generating_2010}, the model does not reflect realistically  how distances scale in optical core networks. 

The Signal-to-Noise Ratio Barabasi Albert (SNR-BA) model was derived to include distance scaling according to the SNR. The model retains aspects from the original BA model, however the probability of two nodes ($i, j \in V$) connecting in a graph with set of nodes $V$ is scaled according to the distance $D(i,j)$ between them \cite{matzner_making_2021}. This probability is given by \eqref{eq:snr_ba_prob}. A weighting $\theta$ is given to how much the distance should influence the probability compared to the degrees ($\delta$) of the nodes. This model retains the sequential generation of a graph, originally in the BA model. This means that nodes are added one by one and then edges are added according to \eqref{eq:snr_ba_prob}. The number of edges can be controlled by limiting the number of edges that are added per node addition.

\begin{equation}
P_\text{SNR-BA}(i,j) = \frac{1}{\left(\displaystyle \sum_{\substack{k \in V}}{\frac{D(i,j)}{D(i,k)}}\right)^\theta}\cdot\frac{\delta_j}{\displaystyle \sum_{\substack{k \in V}}\delta_k}
\label{eq:snr_ba_prob}
\end{equation}

More recently TopoHub has published a set of GG generated topologies ranging between 25 and 500 nodes \cite{jurkiewicz_topohub_2023}. However, as mentioned above, the GG model is deterministic and therefore lacks flexibility. Additionally, some topologies have nodes with degree of 1. This does not represent the minimum resilience requirements of optical core networks \cite{simmons2014optical}. Therefore, we propose to use the SNR-BA model to generate a dataset of synthetic real optical core networks.

\section{Topology Bench: Core Optical Networks}
\label{sec:methods}
\subsection{Dataset of Real World Core Networks}

Topology Bench is a collection of optical network topologies compiled from existing datasets, research literature, and industry reports. Topology Bench primarily focuses on core networks and global networks. The dataset includes a diverse range of network scales, from country-wide to intercontinental networks. The regionality of each network is specified in Appendix A, providing a clear indication of the geographic coverage represented in our dataset. Inclusion criteria were open accessibility, representation of core networks, and provision of geospatial information and node connectivity data linking cities. The dataset is available at \href{https://zenodo.org/records/13921775}{Zenodo}, with a full topology list in Appendix \ref{appendix:A}.

Topology Bench is different from TopoHub, Internet Topology Zoo and SNDLib in that our topology data is homogeneous. The dataset has consistent formatting of node and edge attributes across all topologies, each representing a real-world optical network. 
The data sources and processing methodology for Topology Bench includes the following steps. 
SNDLib topologies were filtered to include only those with real spatial coordinates, excluding 17 networks with abstract coordinates (Figure \ref{fig:limit_data}a). Due to the inaccessibility of the Internet Topology Zoo website, a subset of its topologies was accessed via TopoHub. From TopoHub, we selectively included optical network topologies while omitting IP-based networks. Through an extensive review, we reconstructed 39 additional real optical network topologies, reverse-engineering them into graphs with distinct nodes, edges, and approximate locations. Our synthesis expands the size of publicly available dataset by 61.5\%.

The final dataset of Topology Bench includes an extensive and diverse collection of 105 homogeneous and complete real world optical core network topologies, each represented as a graph with consistent features across the dataset. 
The dataset is characterised by the following features.
\begin{itemize}[noitemsep]
    \item \textbf{\textit{Uniformity:}} Consistency of data (nodes, edges) and tabular format across all topologies. 
    \item \textbf{\textit{Completeness:}} All topologies have undergone rigorous preprocessing to address missing or unknown values. 
    \item \textbf{\textit{Extensivity:}} All real topologies were sourced from standard datasets ~\cite{sdnlib, knight_internet_2011}, research papers~\cite{CETINKAYA201495,simmons2014optical,Lee:15,Korotky:04,9013523,7041849,retail_india_ref} (more than 12\% of real data set), and archived websites~\cite{red_clara, ref_nets, sanet_net} (more than 25\% of real data set). Refer to Appendix \ref{appendix:A} for details. 
    \item \textbf{\textit{Diversity:}} Real topologies are diverse in graph size and geographic and temporal range (see appendix \ref{appendix:A}).
\end{itemize}
\subsubsection{Nodes Data}
As shown in Figure~\ref{fig:topology_bench_tables}, each node in the graph represents a city or Point of Presence (PoP) in the optical network. The characteristics of each node are defined as follows:
\begin{description}[
  leftmargin=2.5cm, 
  labelwidth=2.5cm, 
  font=\normalfont\bfseries, 
  noitemsep, 
  parsep=0pt
]
  \item[Node Identifier] Integer
  \item[Latitude ($\phi$)] Decimal degrees
  \item[Longitude ($\sigma$)] Decimal degrees
  \item[Location] String
  \item[Country] String
\end{description}
The process of obtaining geographic information for each node in the investigated topologies varies depending on the source format of the topology data. 
\begin{itemize}[noitemsep]
    \item \textbf{Case 1:} Latitude and longitude coordinates are available. Reverse geocoding is performed using the GeoPy library to determine the corresponding location details \cite{geopy}.
    \item \textbf{Case 2:} Only node place names are available. The Google Maps API is utilised to obtain the latitude and longitude coordinates for each node.
    \item \textbf{Case 3:} Neither latitude and longitude coordinates nor location names are available. Physical connectivity alone is deemed insufficient for inclusion in the dataset. Topologies in this category are beyond the scope of this study.
\end{itemize}

\begin{figure*}[t]
    \centering
\begin{minipage}[b]{0.45\textwidth}
        \centering
        \scriptsize  \begin{tabular}{|c|c|c|c|c|}
            \hline
            \textbf{Node\_ID} & \textbf{Latitude} & \textbf{Longitude} & \textbf{Location Name} & \textbf{Country} \\ 
            \hline
            1 & 48.2091 & 16.3729 & Vienna & Austria \\ 
            2 & 50.8469 & 4.3518 & Brussels & Belgium \\ 
            3 & 46.2038 & 6.1399 & Geneve & Switzerland \\ 
            4 & 50.0785 & 14.4423 & Karlin & Czechia \\ 
            5 & 50.1122 & 8.6842 & Frankfurt am Main & Germany \\ 
\hline
            
        \end{tabular}
    \end{minipage}
    \hspace{0.03\textwidth} \begin{minipage}[b]{0.45\textwidth}
        \centering
        \scriptsize  \begin{tabular}{|c|c|c|c|}
            \hline
            \textbf{Edge\_ID} & \textbf{Source} & \textbf{Destination} & \textbf{Computed Length (km)} \\ 
            \hline
            1 & 1 & 3 & 1205.75 \\ 
            2 & 1 & 5 & 896.16 \\ 
            3 & 1 & 10 & 326.79 \\ 
            4 & 1 & 16 & 8494.16 \\ 
            5 & 1 & 20 & 416.20 \\ 
\hline
\end{tabular}
    \end{minipage}
    \caption{
    Topology Bench includes tabular datasets for each topology, with 105 real-world topologies and 270,900 synthetic topologies. 
    This is an example of the GEANT topology, presented in an easy-to-view format and ready-to-use for further analysis.
}
    \label{fig:topology_bench_tables}
\end{figure*}

\subsubsection{Edges Data}
Edges represent the fiber-optic connections between nodes. As shown in Figure~\ref{fig:topology_bench_tables}, the characteristics of each edge are as follows:
\begin{description}[
  leftmargin=4.4cm, 
  labelwidth=4.4cm, 
  font=\normalfont\bfseries, 
  noitemsep, 
  parsep=0pt
]
  \item[Edge Identifier] Integer
  \item[Source Node Identifier] Integer
  \item[Destination Node Identifier] Integer
  \item[Fiber Length] Decimal distance in kilometers
\end{description}

The fibre distance was the edge feature considered in this work. Fibre distance affects the throughput of optical networks, ultimately determining resources, given a traffic demand matrix, which includes the number of transponders and wavelengths. Additionally, fibre distance data is readily available for the networked data collected. The length of the fibre corresponding to each edge in the optical network topology, $D_{fibre}$, is expressed in kilometers and calculated as follows:
\begin{equation}
    D_{fibre} = 
    \begin{cases}
    1.5 \cdot D_{hav}  & \quad \text{if $D_{hav} < 1000 \: km$} \\
    1500 \: km & \quad \text{if $1000 \: km \leq D_{hav} \leq 1200 \: km$} \\
    1.25 \cdot D_{hav} & \quad \text{if $D_{hav} > 1200 \: km$}
\end{cases}
    \label{eq:fiber_distance}
\end{equation}
where $D_{hav}$ is the Haversine distance between the two nodes at the extremes of the edge along the sphere's surface. The actual fibre length exceeds the Haversine distance due to factors such as terrain, infrastructure constraints, and installation practices \cite{de_maesschalck_network_1998}. The Haversine distance is computed as: 
\begin{equation}
D_{hav} = 2R \cdot \arcsin(\sqrt{\kappa})
\end{equation}
Where $R$ is the radius of the Earth (approximately $R$ = 6,371 km) and  $\kappa$ is defined in \eqref{eq:sqrt_hav}, where $\phi_1, \phi_2$ are the latitudes of points 1 and 2 in radians respectively and $\sigma_1, \sigma_2$ are the longitudes of points 1 and 2 in radians respectively.
\begin{equation}
\kappa = 
\sin^2 \left(\frac{\phi_2-\phi_1}{2}\right) 
 +\cos(\phi_1)\cdot\cos(\phi_2)\sin^2\left(\frac{\sigma_2 - \sigma_1}{2}\right) 
\label{eq:sqrt_hav}
\end{equation}
\subsubsection{Topology Representation and Visualisation}
To ensure reproducibility and standardisation in the visualisation of optical core networks, we implemented a systematic approach for topology representation.
\begin{itemize}[noitemsep]
\item \textbf{Data Standardisation:} All optical core network topologies were transformed into a uniform, human- and machine-readable format.  Each network topology is defined by two distinct comma-separated value (.csv) files containing node-level and edge-level data as \textbf{ \texttt{nodes\_<topologyname>.csv}} and \textbf{\texttt{edges\_<topologyname>.csv}}, respectively. 
\item \textbf{Graph Construction:} The NetworkX library~\cite{NetworkX} was used to create a graph representation of each topology, supporting algorithms, metric calculations, and visualisations.
\item \textbf{Geospatial Mapping:} Using NetworkX and the Folium library \cite{folium}, nodes were placed based on latitude and longitude from \texttt{nodes\_<topologyname>.csv}, and edges were connected using data from \texttt{edges\_<topologyname>.csv}.
\end{itemize}
To complement our real optical network dataset, we additionally generate synthetic networks using the SNR-BA graph model.

\subsection{Dataset of Synthetic Optical Core Networks}
Two datasets were generated, a small and large synthetic network dataset made of 900 and 270,000 networks respectively. For both datasets, node and edge generation are as follows.
\begin{itemize}[noitemsep]
    \item \textbf{Node generation}. The network size is an integer $n \in \mathbb{N}: 10\le n \le 100$. 
    Most collected real topologies (98\%) had a size below or equal to 100 nodes. Although core topologies may have a larger number of nodes, the network sizes in the current version of Topology Bench were restricted to below or equal to 100 nodes to align with the published data. Larger network sizes will be left for future work and the authors would be happy to receive additional topology data.
    Nodes are located randomly in a rectangle defined by 4 coordinate pairs of longitude and latitude. New nodes added to the topology cannot be closer to an existing node than a pre-defined radius of $R=80$~km. Three rectangles are defined for large, medium, and smaller distance scales, as defined in Table \ref{tab:coordinates}. The largest distance scale is defined over coordinates that span the North American continent, where the largest real world networks are found. The medium distance scale, defines a rectangle half the size of the largest one. This approximately corresponds to the European continent scale. The smallest distance scale, again halves the latitudes used for the medium ones.
    \item \textbf{Edge generation}. The number of edges is defined as $m=d~\cdot~n$, given that within real optical networks the number of edges scales approximately linearly with the number of nodes, as investigated later in Section ~\ref{sec:synthetic_network_analysis}.  Edges are generated using the SNR-BA model described in Section ~\ref{sec:related_work}\ref{sec:existing_synthetic_work}, where a certain number of edges are added with each node addition ~\cite{matzner_making_2021}. The value for the distance weighting constant $\theta=5$   was used for the generation of these networks. This value was optimised previously in \cite{matzner_making_2021} to give network structures closest to those in real networks.
    About 93\% of the 105 real networks collected are planar graphs. Therefore, network planarity is an important feature of real networks. The SNR-BA network does not ensure planarity. To include planarity in the generation of SNR-BA networks would significantly restrict edge choices – influencing the final structures of the resultant networks. In addition, if considering planarity as a hard constraint in the generation of networks, then the network densities of networks are restricted, as higher   network densities are not achievable ($\frac{6n-12}{n(n-1)}$) with planarity as a constraint. Therefore, planarity is not considered as a hard constraint, however 48\% of the synthetic-small networks were, indeed, planar graphs.
\end{itemize}

The small synthetic dataset has 10 networks per network size, 90 different network sizes (uniformly spread across $n=[10, 100, 1]$), a single network density ($d=1.2$) and a single distance scale (large). The large synthetic dataset encompasses 100 networks per network size, 90 network sizes, 10 different network density values ($d=[1.2, 4.8, 0.4]$, where the last number denotes the step size) and 3 distance scales. This large range of network density values is generated to enable researchers to evaluate not only the cases of “realistic network density” values but also evaluate the impact of denser topologies for future network growth. Additionally, machine learning applications require varying levels of network density for generalisation.
\begin{table}
    \centering
    \resizebox{\columnwidth}{!}{
    \begin{tabular}{|c|c|c|c|c|c|c|}
    \hline
         & $l_1$ & $l_2$ & $m_1$ & $m_2$ & $s_1$ & $s_2$  \\
        \hline
        lon & -124.86 & -67.39 & -124.86 & -96.12 & -124.86 & -96.12 \\
        \hline
        lat & 31.68 & 48.1261 & 31.68 & 48.12 & 31.68 & 39.90 \\
        \hline
    \end{tabular}
    }
    \caption{Coordinates for large distance scales ($l_1, l_2$), medium distance scales ($m_1, m_2$) and small distance scales ($s_1, s_2$). }
    \label{tab:coordinates}
\end{table}

\subsection{Network Analysis Metrics}

\begin{table*}[h!]
\centering
\begin{tabular}{| m{3cm} | m{7.5cm} | m{7cm} |}
\hline
\textbf{Metric} & \textbf{Definition} & \textbf{Formulation} \\
\hline

Node Degree ($d(u)$) & The degree of node $u$ is the number of edges connected to it. The average value is calculated by $\bar{d} = \frac{2m}{n}$ & 
\begin{equation}
    d(v) = \sum_{u \in V}A(v, u)
    \label{eq:degree}
\end{equation}
\\
\hline
Diameter (\( D \)) & The greatest distance between any pair of nodes in the graph. Here 
$d(u,v)$ is the shortest path distance between nodes $u$ and $v$. & 
\begin{equation} 
    D = \max_{u,v \in V} d(u, v)
    \label{eq:diameter}
\end{equation}
 \\ 
\hline
Network Density (\( ND \)) & A measure of how many edges are in the graph compared to the maximum possible number of edges. & 
\begin{equation}
    ND = \frac{2m}{n(n-1)}
    \label{eq:edge_density}
\end{equation}
\\

\hline
Average Shortest Path Length (\( L \)) & The average number of hops along the shortest paths for all possible pairs of nodes. & 
\begin{equation}
 L = \frac{1}{n(n-1)} \sum_{u \neq v \in V} d(u, v)  
 \label{eq:average_shortest_path}
\end{equation}
\\    
\hline

Clustering Coefficient (\( C(v) \)) & The probability that nodes in the neighbourhood of $v$ ($\mathcal{N}(v)$) share same neighbours of $v$. The average value over all nodes in a graph is referred to as $C_{avg}$. & 
\begin{equation}
C(v)=\frac{2 \sum_{i \in \mathcal{N}(v)}\sum_{j \in \mathcal{N}(v)\setminus i} A(i,j)}{d(v)(d(v)-1)} \quad \forall d(v) \geq 2
\label{eq:clustering_coefficient}
\end{equation}
\\
\hline
Maximum Edge Betweenness (\( C_{EB}^{\max} \)) & The largest number of paths crossing over an edge in a graph. Here $\sigma_{s t}$ is the number of shortest paths from $s$ to $t$ and $\sigma_{s t}(e)$ is the number of those paths passing through $e$.& 

\begin{equation}
    C_B^{\max} = \max_{e \in E} \bigg ( \sum_{s \neq v \neq t \in V} \frac{\sigma_{st}(e)}{\sigma_{st}} \bigg )
    \label{eq:edge_betweeness}
\end{equation}
 \\
\hline
Maximum Node Betweenness (\( C_{NB}^{\max} \)) & The largest number of paths crossing over a node in a grap. Here $\sigma_{s t}$ is the number of shortest paths from $s$ to $t$ and $\sigma_{s t}(v)$ is the number of those paths passing through $v$.& 

\begin{equation}
    C_B^{\max} = \max_{v \in V} \bigg (\sum_{s \neq v \neq t \in V} \frac{\sigma_{st}(v)}{\sigma_{st}} \bigg )
    \label{eq:node_betweeness}
\end{equation}
 \\

\hline
Global Efficiency (\( E_{\text{glob}} \)) & A measure of the average efficiency of the network in exchanging information. & 

\begin{equation}
    E_{\text{glob}} = \frac{1}{n(n-1)} \sum_{u \neq v \in V} \frac{1}{d(u,v)}
    \label{eq:global_efficiency}
\end{equation}
 \\
\hline
Spectral Radius (\( \rho(A) \)) & The largest absolute value of the eigenvalues of adjacency matrix $A$. & 

\begin{equation}
    \rho(A) = \lambda_V
    \label{eq:spectral_radius}
\end{equation}
 \\
\hline
Algebraic Connectivity (\( \lambda_{AC} \)) & The second smallest eigenvalue of the Laplacian matrix. & 
\begin{equation}
    \lambda_{AC} = \lambda^{L}_1
    \label{eq:algebraic_conn}
\end{equation}
\\
\hline
Weighted spectral distribution ($WSD(G)$) & The weighted sum of the eigenvalue distribution of the normalised Laplacian & \begin{equation}
    WSD(G) = \sum_{k \in K} (1-k)^N f(\lambda^{L_D} = k)
    \label{eq:WeightedSpectralDistribution}
\end{equation}
\\
\hline
\end{tabular}
\caption{
Definitions of metrics used for network topology statistics, correlations, comparisons, and topology selection.
}
\label{table:graphmetrics}
\end{table*}
The final performance of optical networks (throughput, resilience, latency) depends on both the technology used (number of transponders, type of transponder, modulation, amplifiers, type of fibre) and the mathematical properties of the network. Therefore, the graph theoretical properties of optical networks and the technology used within them are closely linked. The spatial properties within the network might determine what modulation is possible, how many amplifiers or which amplifiers to use. Therefore, it is important to test on a range of networks with varying graph properties.
The topology analysis metrics used in this paper are defined in Table \ref{table:graphmetrics}. 
Graph metrics are considered structural if they ignore fibre distances, or spatial if they incorporate fibre distances.
For the graph metrics, we denote a graph as $G(V, E)$ with set of nodes $V$ and set of edges $E$. The number of nodes and edges are referred to as $n=|V|$ and $m=|E|$, respectively. $G$ can be fully encoded in the adjacency matrix $A(i,j)$ as defined in \eqref{eq:AdjacencyMatrix}. 

\begin{equation} A(i, j) = 
  \begin{cases}
    1  & \quad \text{if node $i$ is connected to node $j$} \\ 
    0  & \quad \text{otherwise} 
  \end{cases}
  \label{eq:AdjacencyMatrix}
\end{equation}

Any network metric that is derived from the spectrum of the Adjacency, Laplacian or normalised Laplacian (defined in \eqref{eq:Laplacian} and \eqref{eq:NormalisedLaplacian}) is referred to as a spectral metric. Spectral metrics originate from spectral graph theory \cite{cetinkaya_comparative_2014} and generalise structural analysis over graphs of different sizes. For the first spectral metric (Spectral Radius) we denote the set of real eigenvalues of adjacency matrix $A$ by $\lambda(A) = \{\lambda_0, \lambda_1, ..., \lambda_V\}$.
Spectral methods can also be defined over the distribution of eigenvalues from the Laplacian matrix of the network defined as in \eqref{eq:Laplacian}, where $D_L$ is the degree matrix, defined in \eqref{eq:DegreeMatrix}. $L$ is a real symmetric matrix with real eigenvalues $\lambda(L) = \{\lambda^{L}_0, \lambda^{L}_1, ..., \lambda^{L}_V\}$.

\begin{equation}
    L = D_L - A
    \label{eq:Laplacian}
\end{equation}
\begin{equation}
D^{i,j}_L = 
\begin{cases}
    d(i)  & \quad \text{if i = j} \\ 
    0  & \quad \text{otherwise} 
  \end{cases}
  \label{eq:DegreeMatrix}
\end{equation}
From the Laplacian matrix we can define the metrics algebraic connectivity and the spectral gap. The algebraic connectivity can be considered a measure of how bottle-necked a network is \cite{fan_chung_spectral_2006}, whereas the spectral gap has been shown to relate to the robustness of networks \cite{yazdani_resilience_2011}.

To further generalise the analysis the Laplacian matrix is normalised, resulting in the normalised Laplacian matrix of the graph, defined in \eqref{eq:NormalisedLaplacian}. $L_D$ is a real symmetric matrix with real eigenvalues $\lambda(L_D) = \{\lambda^{L_D}_0, \lambda^{L_D}_1, ..., \lambda^{L_D}_V\}$.

\begin{equation}
    L_D =  D_L^{-\frac{1}{2}}LD_L^{-\frac{1}{2}} = I - D_L^{-\frac{1}{2}}AD_L^{-\frac{1}{2}}
    \label{eq:NormalisedLaplacian}
\end{equation}

$L_D$ allows to define the weighted spectral distribution (WSD) metric \cite{fay_weighted_2008}, which summarises the structure of any graph. Given the eigenvalues of $L_D$ one can construct the WSD as a sum over the probability function defined in \eqref{eq:WeightedSpectralDistribution} in Table \ref{table:graphmetrics}. Here $k \in K$ are the bins used to bin the eigenvalue distribution. The function $f(\lambda = k)$ is the fraction of eigenvalues falling into bin $k$. The constants $K=40$ and $N=4$ are chosen according to \cite{fay_weighted_2008}. This metric describes the spectral distribution of a graph, giving unique values for specific structures in graphs.

We calculated all the graph metrics in Table \ref{table:graphmetrics} for all real topologies and saved them in two types of files. The first is a consolidated file containing metrics for all topologies, enabling comparative analysis. The second consists of individual files for each topology, allowing detailed examination of specific networks. Both are available in the \hyperlink{https://github.com/TopologyBench/analysis}{GitHub} analysis repository.

The metrics in Table \ref{table:graphmetrics} are applied in Sections \ref{sec:real_network_analysis} and \ref{sec:synthetic_network_analysis} to explore the structural, spatial, and spectral properties of both real and synthetic networks. A subset of these metrics are later used in Section \ref{sec:topology_selection} for clustering and topology selection.
\section{Analysis of Real Networks}
\label{sec:real_network_analysis}
The following sub-sections investigate the structural, spectral and spatial properties of the collated real network dataset.

\subsection{Dataset Trends}
\begin{table*}[h!]
\centering
\begin{tabular}{|l|c|c|c|c|}
\hline
\textbf{Property} & \textbf{Mean} & \textbf{Standard Deviation} & \textbf{Minimum} & \textbf{Maximum} \\ \hline
Number of Nodes & 26.028571 & 20.285205 & 6.000000 & 142.000000 \\ \hline
Number of Edges & 34.209524 & 29.252119 & 6.000000 & 180.000000 \\ \hline
Average Node Degree & 2.546325 & 0.431357 & 1.750000 & 3.894737 \\ \hline
Average Clustering Coefficient & 0.113068 & 0.136595 & 0.000000 & 0.686111 \\ \hline
Network Density & 0.151694 & 0.098991 & 0.017980 & 0.476190 \\ \hline
Maximum Node Betweenness Centrality & 0.447834 & 0.137315 & 0.200000 & 0.897233 \\ \hline
Average Node Betweenness Centrality & 0.128383 & 0.047615 & 0.056304 & 0.291071 \\ \hline
Maximum Edge Betweenness Centrality & 0.305709 & 0.091543 & 0.132103 & 0.545455 \\ \hline
Average Edge Betweenness Centrality & 0.141382 & 0.070080 & 0.038116 & 0.317188 \\ \hline
Diameter (Hops) & 8.047619 & 4.105299 & 2.000000 & 28.000000 \\ \hline
Average Shortest Path Length (Hops) & 3.590185 & 1.380766 & 1.523810 & 9.818400 \\ \hline
Global Efficiency & 0.413816 & 0.120698 & 0.153517 & 0.738095 \\ \hline
Normalized Spectral Radius & 0.642123 & 0.137651 & 0.331924 & 1.000000 \\ \hline
Normalized Algebraic Connectivity  & 0.105178 & 0.115573 & 0.005518 & 0.730726 \\ \hline
Normalized Weighted Spectral Distribution & 0.241183 & 0.067606 & 0.092094 & 0.431949 \\ \hline
Average Shortest Path Length (km) & 2118.983788 & 1886.598290 & 165.431471 & 10882.072506 \\ \hline
Absolute Average Link Length (km) & 687.918551 & 688.937704 & 32.527273 & 4159.836963 \\ \hline
Absolute Standard Deviation of Shortest Path Lengths (km) & 1310.314661 & 1330.498934 & 91.530714 & 7970.808101 \\ \hline
Absolute Diameter (km) & 5343.295654 & 5459.772472 & 349.500000 & 36855.121523 \\ \hline
\end{tabular}
\caption{Summary statistics of the structural, spectral, and spatial properties of 105 real-world optical networks.}
\label{tab:network_stats}
\end{table*}

Table \ref{tab:network_stats} summarises key graph, spectral, and spatial metrics (in km) for the real network dataset, including mean, standard deviation, minimum, and maximum values. Notable trends:

\begin{itemize}[noitemsep]
    \item \textbf{Diverse network sizes influence density and sparsity:} The range of node counts (6 to 142) and edges (6 to 180) demonstrates a diverse set of networks. The low average node degree (2.55), average clustering coefficient (0.113) and network density (0.152) strongly indicate that these are predominantly sparse networks. It reflects the cost-conscious nature of optical network deployment, where each link represents significant infrastructure investment. The high variability in diameter (2 to 28 hops) indicates a mixture of network structures.
    \item \textbf{Topological patterns suggest centralized designs:} 
    Our analysis revealed that 40 real optical networks have an average clustering coefficient of zero, indicating tree-like, star-like, or grid-like structures with no local clustering. This contrasts with random graphs, where some clustering is expected. In optical networks, this suggests a centralized design with hub nodes or linear/ring topologies, common in long-haul networks.
    \item \textbf{Shorter paths improve efficiency:} 
    The low average shortest path length indicates that these networks are structurally efficient. Despite the sparsity observed, the networks maintain relatively short paths between nodes. The low average path length, combined with the earlier observation of low clustering, suggests that many of these networks exhibit small-world properties known for efficient information transfer. This results in lower network latency and efficient data routing through the network, reducing the load on individual links. It suggests that high-degree nodes were strategically placed while designing the network. 
    \item \textbf{Efficiency as an indicator of performance:} Global efficiency has been shown to correlate with network throughput \cite{higashimori_impact_2022}. An average global efficiency (defined in \eqref{eq:global_efficiency}) of 0.41 for 105 networks, suggests moderate overall network performance in terms of network throughput. Highest efficiency is seen in NETRAIL (0.738) followed by EPOCH and LAYER42 (0.711). The standard deviation of 0.12 suggests a moderate level of variability in efficiency across the dataset. 
    \item \textbf{Critical nodes and links dictate network hierarchy:} 
     The average maximum node betweenness centrality (mean 0.448, standard deviation 0.137) indicates the presence of critical nodes in most networks, with a large gap between the maximum (0.897) and average values, suggesting a hierarchical structure. Similarly, the average maximum edge betweenness centrality (mean 0.306, stdev 0.091) highlights the presence of critical links. The lower mean for edges compared to nodes suggests that critical nodes play a more important role than edges in network connectivity.
     \item \textbf{Spectral properties reveal network connectivity strengths and vulnerabilities:} 
     A high normalized spectral radius suggests a well-connected network where information or data can flow efficiently where a low algebraic connectivity indicates a network with weak points or bottlenecks, where the overall connectivity relies heavily on a few critical connections. The maximum normalized spectral radius of 1.0 suggests the presence of at least one network with very high connectivity. The minimum normalized algebraic connectivity of 0.0055 indicates the existence of networks with very poor connectivity, possibly with severe bottlenecks or near-disconnected components.
     \item \textbf{Spatial properties highlight design diversity across network scales:} 
     Our dataset covers diverse network scales and types, from small core networks (minimum link length of 32.5 km) to intercontinental networks (maximum link length of 4159.8 km). The set of 105 real networks includes 9 networks with submarine (or subsea) links. Submarine systems span longer distances than terrestrial links, often connecting continents which impacts factors such as latency, signal degradation, and number and spacing of amplifiers. The diameter range further emphasises the diversity, spanning from regional networks (349.5 km) to continental and global networks (36855.1 km). High standard deviations in both link lengths and diameters indicate significant variability in network designs.
\end{itemize}
Finally, we analysed the topologies for planarity and resilience.
\begin{itemize}[noitemsep] 
    \item \textbf{Network Planarity reflects geographical constraints and cost-efficiency:}. Out of 105 topologies, 98 (93.3\%) networks are planar, suggesting that these optical networks have relatively simple topological structures. The prevalence of planarity reflects the geographical constraints in deploying optical networks.
    Planar graphs have a limited number of edges relative to their vertices. For a planar graph with $n$ vertices ($n \geq 3$), the number of edges $m$ is bounded by $m \leq 3n - 6$. This reinforces the trend of sparse connections, a natural design choice as the cost is directly proportional to the length of the fibre. 
    \item \textbf{Bridges reduce network resilience by increasing vulnerability to single-point failures:} More than half, 53 out of 105 of the networks contain bridges. Bridges are edges whose removal would disconnect the graph. Thus, 50.48\% of real-world networks are vulnerable to single-point failures. Bridges also represent a lack of alternative paths between different parts of the network and the limited redundancy can impact the network performance and resilience.
    On the other hand, survivable topologies are defined as networks that are minimally bi-connected. That is, at least two edges must be removed to disconnect the graph, which ensures that no single node or edge failure can disconnect the network. Topology Bench includes 52 such survivable topologies. We observe a clear dichotomy in our dataset between more resilient designs and those vulnerable to single point failures.
    \end{itemize}
\subsection{Correlation Analysis of Network Metrics}
A correlation analysis was conducted to uncover underlying relationships between the computed network metrics. We used the Pearson correlation coefficient, a statistical measure that quantifies the linear relationship between two variables. 
\subsubsection{Correlation Matrix Computation}
\begin{figure*}[h!]
    \centering
\includegraphics[width=\linewidth]{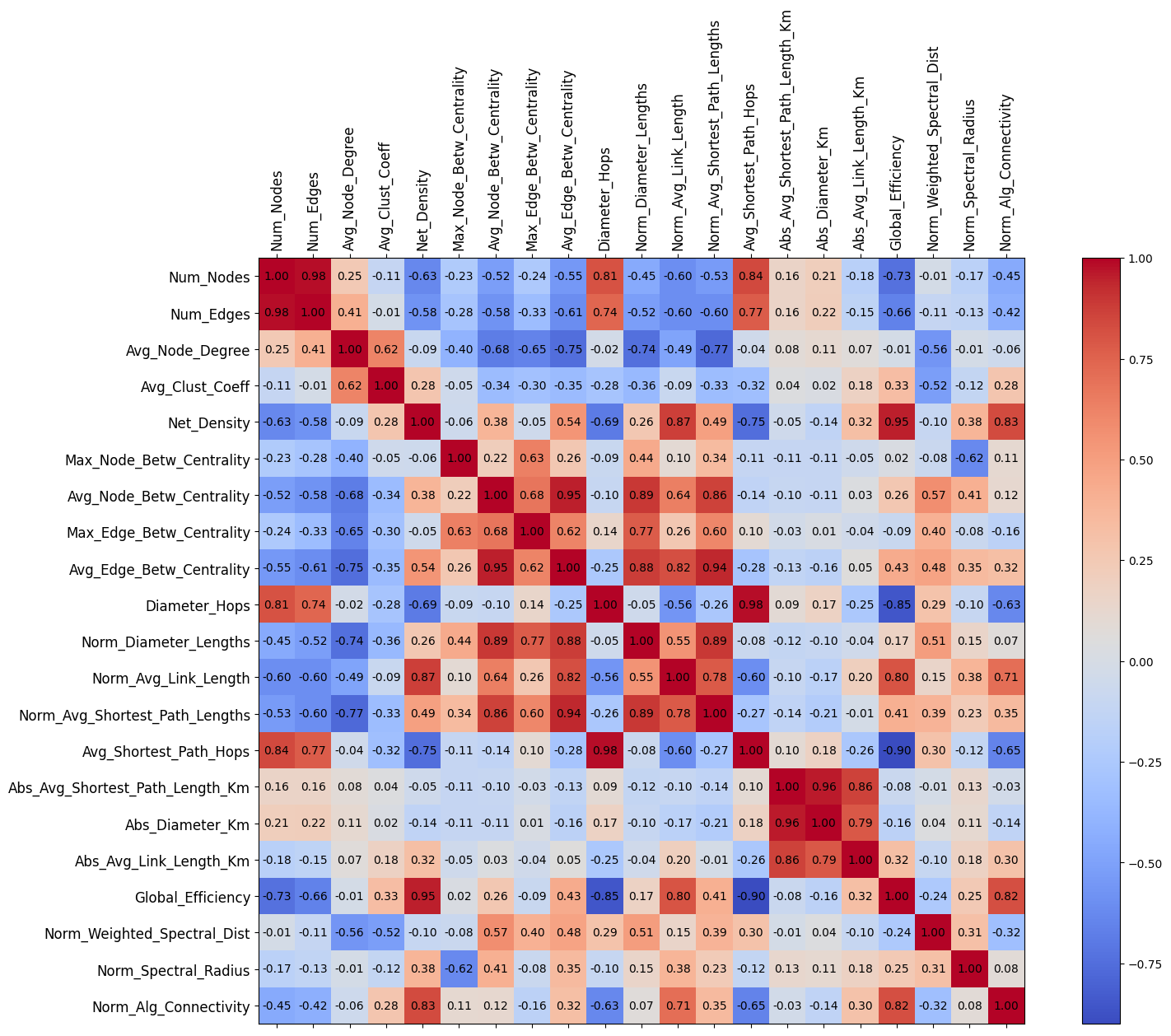}
    \caption{Correlation heatmap of 21 graph metrics for 105 real world optical networks in Topology Bench.}
    \label{fig:heatmap}
\end{figure*}

Following the selection of relevant network metrics, we conducted a correlation analysis to understand the relationships among them. We computed the Pearson correlation coefficient for each pair of metrics $x$ and $y$, $\rho_{x,y}$, as:
$$
\rho_{x,y}=\frac{\sum_{i=1}^q\left(x_i-\bar{x}\right)\left(y_i-\bar{y}\right)}{\sqrt{\sum_{i=1}^q\left(x_i-\bar{x}\right)^2 \sum_{i=1}^q\left(y_i-\bar{y}\right)^2}}
$$
where $x_i$ and $y_i$ are the individual data points, $\bar{x}$ and $\bar{y}$ are their respective means, and $q$ is the number of data points.
Figure \ref{fig:heatmap} shows a heatmap of the correlation matrix, where color coding represents the strength and direction of the correlations. A summary statistic, the average absolute correlation, was calculated as:
\[
AvgCorr = \frac{\sum \mid r_{ij} \mid - n}{n(n-1)}
\]
where $|r_{ij}|$ is the absolute value of the correlation between metrics $i$ and $j$, and $n$ is the number of metrics. The computed average absolute correlation of 0.3469 indicates a moderate level of correlation among the network metrics, suggesting that while some metrics are related, many still capture unique information.
\subsubsection{Implication of Strong Correlations}
We highlight key graph metric correlations in this section
\begin{itemize}[noitemsep]
    \item \textbf{Improving algebraic connectivity enhances  efficiency:} A key insight is the strong positive correlation between Normalised Algebraic Connectivity and Global Efficiency (0.82). Well-connected networks in terms of algebraic connectivity are also efficient in terms of information flow \cite{chatelain_topological_2009}. However, it is important to note that high level of correlation between topology features does not necessarily confirm the same relationship to the overall network performance. Network designers can focus on improving algebraic connectivity as a proxy for enhancing efficiency. 
    \item \textbf{Increasing average node degree reduces network distance:} The negative correlation (-0.77) between Normalised Average Shortest Path Length and Average Node Degree indicates that higher average node degrees lead to shorter average path lengths. For network designers, this means that increasing average node degree is an effective way to reduce overall network distance. Shorter distances result in lower latency and reduced signal degradation.
    \item \textbf{Larger networks need strategic path diversification and redundancy to reduce critical edge reliance:} The strong positive correlations of Average Edge Betweenness Centrality with Normalized Average Shortest Path Length (0.94) and Normalized Diameter (0.88) suggest that networks with longer relative paths tend to have edges that are more critical for traffic flow. This implies that, as networks grow in size or geographical spread (increasing diameter), the reliance on critical edges for traffic flow becomes more pronounced. This underscores the need for strategically adding redundant paths or shortcuts in networks with longer relative paths to reduce dependence on critical edges.
    \item \textbf{Balancing network density and size is important for maintaining efficiency in large-scale optical networks:} The extremely strong correlation (0.95) between Network Density and Global Efficiency underscores the critical role of density in network performance. On the other hand, the negative correlation between Global Efficiency and Network Size highlights the challenge of maintaining efficiency in larger networks, which is a consideration for designing large-scale optical network topologies.
    \item \textbf{Optical networks with higher network density tend to cover larger areas, requiring balanced spatial and topological design:} The strong positive correlation (0.87) between Normalized Average Link Length and Network Density (Edge Density) suggests that networks with higher connectivity tend to have relatively longer links. While Network Density and Normalized Average Node Degree are mathematically equivalent, considering both provides insight into network structure from different perspectives - overall connectivity and individual node connections. This relationship between spatial characteristics (link lengths) and topological metrics (network density and node degree) underscores the need to consider multiple aspects in optical network design, indicating that increased connectivity often comes with extended geographical coverage and higher per-node connection rates. 
     \end{itemize} 
\subsubsection{Limitations and assumptions:}
The Pearson correlation coefficient assumes linearity, homoscedasticity, and absence of significant outliers. It also assumes that variables are continuous and measured on interval or ratio scales. Importantly, while normality is often cited as an assumption, it is primarily required for inference rather than for the calculation of the coefficient itself.
The calculated correlation coefficient itself remains a valid descriptor of the linear association in the sample, even with non-normal data. 
When checking for normality, we found 3 out of 21 variables to have normal distributions. Further, when checking for linearity, we found that 63 out of 210 variable pairs demonstrated linear relationships. Lastly, we found no homogeneity of variance across the dataset, indicating heteroscedasticity.
Despite violations of normality and homoscedasticity assumptions, our substantial sample size of 105 real world optical networks enhances the robustness of Pearson's correlation results. This larger sample size, well above the conventional threshold of 30 for statistical inference, mitigates the impact of non-normality and outliers, strengthening the reliability of our findings.
\subsubsection{Outlier Detection}
The z-score for each feature $x$ is computed as:

\begin{equation}
  z_i=\frac{x_i-\mu_x}{\sigma_x}
  \label{eq:z_score}
\end{equation}

A data point is flagged as an outlier if $\left|z_i\right| > 3$. The maximum absolute z-score is used to identify the most significant outlier feature, and the total number of features contributing to outlier status is counted. Our analysis identified 19 optical networks as outliers, with up to 4 metrics contributing to each case. Notably, 10 were flagged based on a single feature, likely representing unique design choices. These topologies are documented in Appendix \ref{appendix:A}. Given that most metrics for these networks fall within normal ranges, and their potential significance in understanding diverse optical network architectures, we recommend retaining these topologies in our analysis.

\section{Analysis of Synthetic Networks}
\label{sec:synthetic_network_analysis}
To validate the SNR-BA model for generating synthetic network topologies, we need to compare its output with real topologies. However, the model's geometric nature means network structure depends on node positioning. To ensure fair comparison, we generated an additional set of synthetic networks using the exact node positions of real networks. We selected 52 bi-connected real networks (as SNR-BA focuses on survivable networks) and generated 100 SNR-BA networks for each, totaling 5200 networks. This approach allows for a more accurate validation of the SNR-BA model by aligning synthetic network generation with real-world node distributions.

\subsection{SNR-BA Model Validation}
\label{sec:snr_ba_model_validation}

\begin{figure*}[h!]
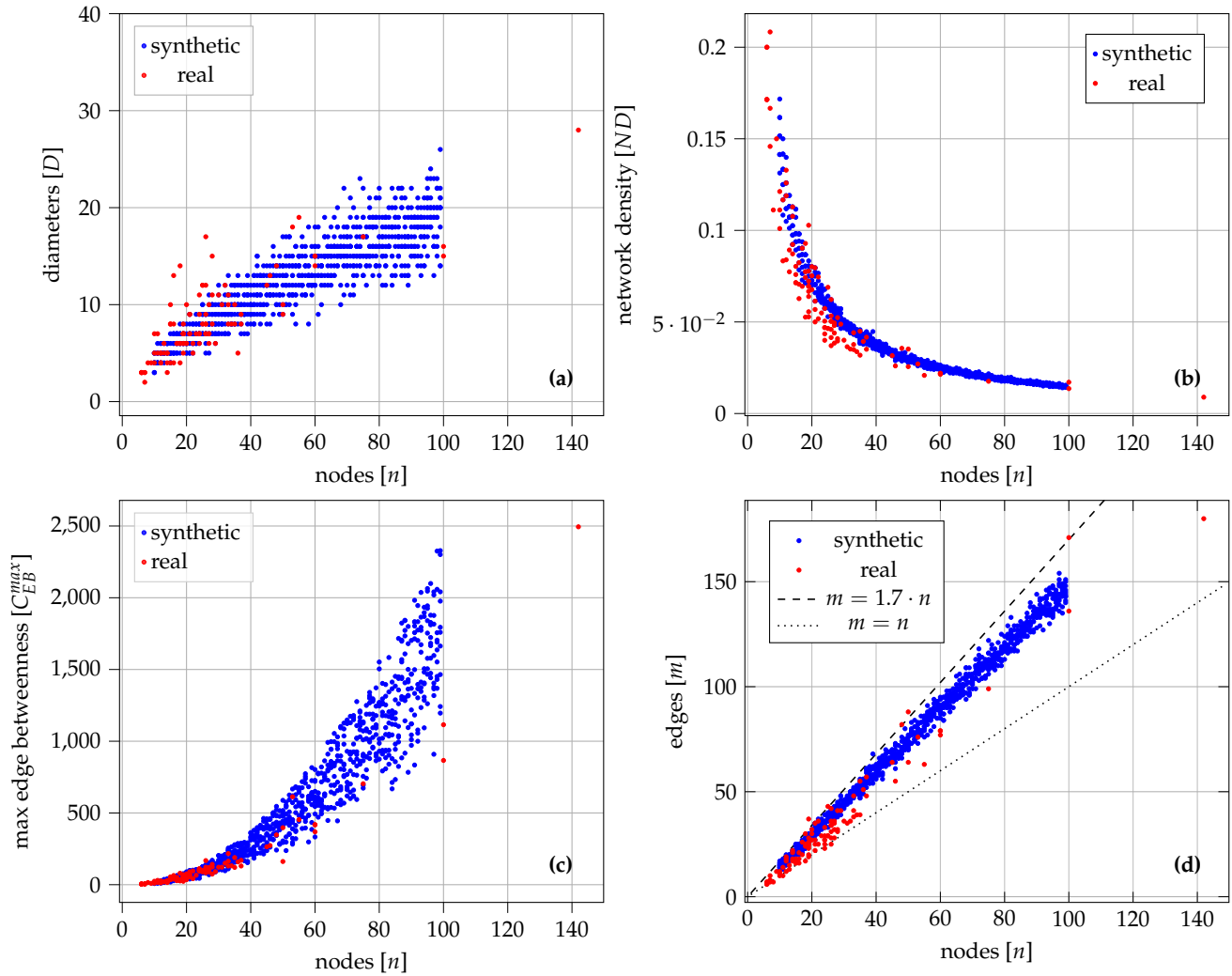

    \clearpage{}\definecolor{darkgray176}{RGB}{176,176,176}
\definecolor{lightgray204}{RGB}{204,204,204}

\clearpage{}
    \caption{ Distribution of (a) the diameter (hops), (b) network density, (c) maximum edge betweenness and number of edges (d)  as a function of the number of nodes for small synthetic and real optical network datasets.}
    \label{fig:synthetic_small_analysis}
\end{figure*}

\begin{table}[h!]
    \centering
    \begin{tabular}{|l|c|}
        \hline
         \textbf{Network Property} & $\boldsymbol{p_{KS}}$\\
         \hline
         Degree ($n \leq 40$)& 0.40\\
         Degree ($n > 40$)& 0.001 \\
         Diameter & 0.99\\
         Normalised Laplacian & 0.15\\
         Fibre Lengths & 0.27\\
         \hline
    \end{tabular}
    \caption{Summary of Kolmogorov-Smirnov 2 sample test statistics ($p_{KS}$) for degree, diameter, normalised Laplacian eigenvalue and fibre lengths distributions. The requirement is that $p_{KS} \geq 0.05$ for two sample sets to originate from the same distribution.}
    \label{tab:snr_ba_val}
\end{table}

The Kolmogorov-Smirnov 2 sample test (KS2, $p_{KS}\geq$ 0.05 threshold) \cite{wang2003} is calculated between the 52 real survivable networks and 5200 synthetically generated ones for degree, diameter, spectral, and fibre distance distributions and summarised in Table \ref{tab:snr_ba_val}. The KS2 allows for comparison of two sets of samples and to determine if they derive from the same distribution.

Degree distributions matched well for networks with $n \leq 40$ ($p_{KS} = 0.40$), but not for larger networks ($p_{KS} = 0.0017$), for which the model produced more high-degree nodes than real networks.

The distributions of network diameter between the 52 real and 5200 synthetically generated networks have excellent agreement ($p_{KS}$ = 0.99), suggesting the SNR-BA model replicates the real distribution, which is significant given the diameter's operational importance in optical networking.

Spectral properties of networks were investigated using normalised Laplacian eigenvalues. These eigenvalues ($\lambda^{L_D}_i$) lie in the range from 0 to 2. Zero eigenvalues denote connected components. The second smallest eigenvalue (algebraic connectivity) correlates to network connectivity \cite{cetinkaya_comparative_2014}. The KS2-test ( $p_{KS}= 0.15 \geq 0.05$) suggests similar eigenvalue distributions and structures between real and synthetic networks, suggesting similar overall network structures.

To complete the analysis, fibre lengths are investigated. The KS2-test yielded a $p_{KS}$ value of 0.27, suggesting that fibre length distributions likely come from the same distribution. This indicates that the SNR-BA model produces realistic spatial properties similar to real survivable networks.

\subsection{Small Synthetic Network Dataset (900 Networks)}
\label{sec:small_synthetic_network_dataset}

\begin{figure}[b!]
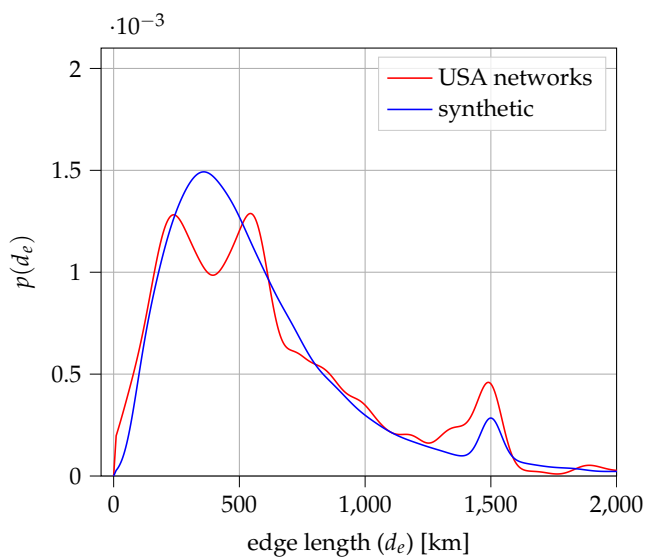

\centering
\clearpage{}


\clearpage{}
\caption{Distance distribution of 16 North American networks compared to those of small synthetic optical networks.}
\label{fig:usa_network_distances}
\end{figure}

\begin{figure*}[t!]
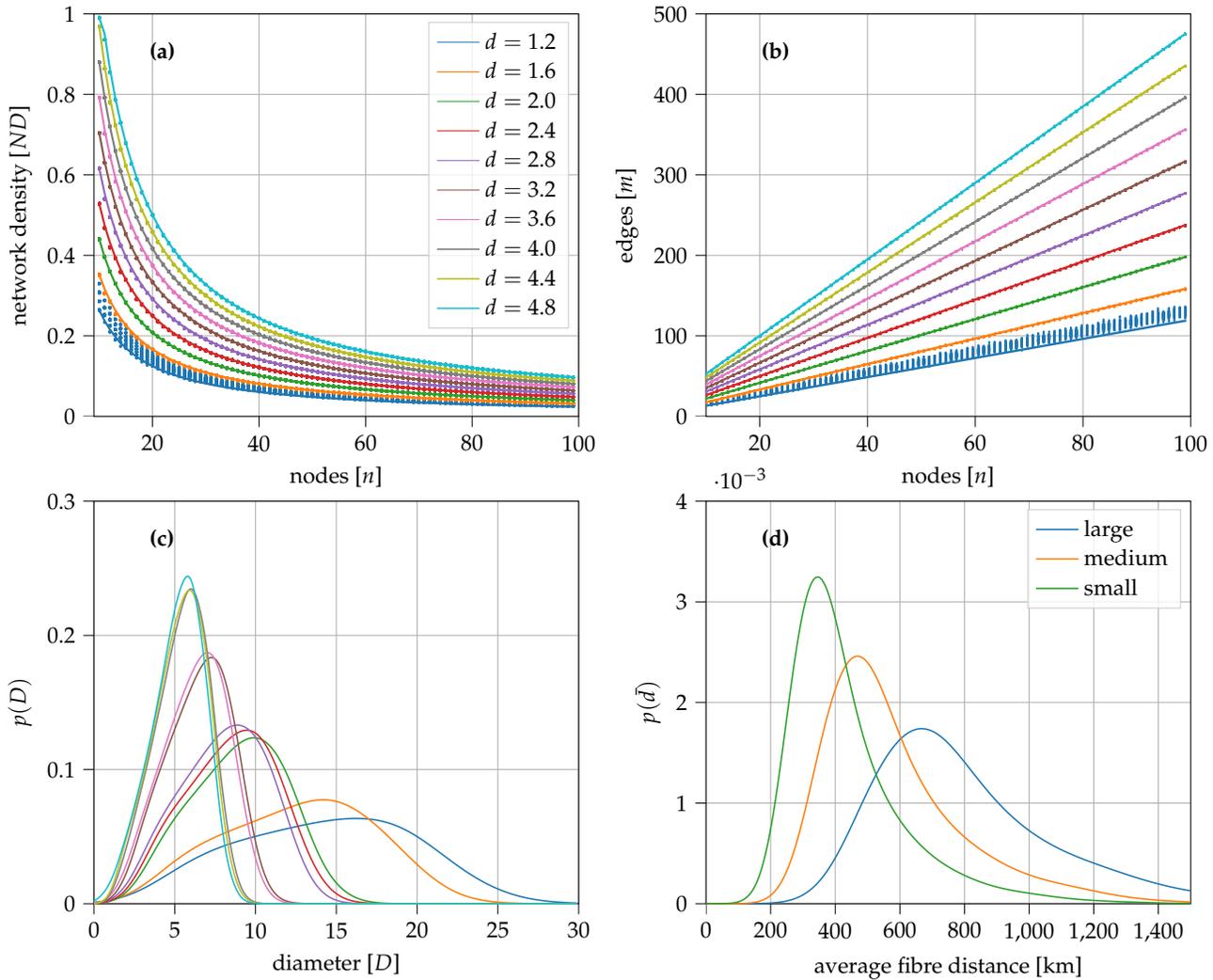

    \definecolor{crimson2143940}{RGB}{214,39,40}
\definecolor{darkgray176}{RGB}{176,176,176}
\definecolor{darkorange25512714}{RGB}{255,127,14}
\definecolor{darkturquoise23190207}{RGB}{23,190,207}
\definecolor{forestgreen4416044}{RGB}{44,160,44}
\definecolor{goldenrod18818934}{RGB}{188,189,34}
\definecolor{gray127}{RGB}{127,127,127}
\definecolor{lightgray204}{RGB}{204,204,204}
\definecolor{mediumpurple148103189}{RGB}{148,103,189}
\definecolor{orchid227119194}{RGB}{227,119,194}
\definecolor{sienna1408675}{RGB}{140,86,75}
\definecolor{steelblue31119180}{RGB}{31,119,180}

     \caption{(a) Network density of large synthetic dataset with $m = d \cdot n$. (b)  Number of edges for the large synthetic dataset with different values of $d$. (c) Probability distribution of diameters of large synthetic dataset. (d) Probability distribution of fibre lengths in large, medium, and small distance scales of the large synthetic dataset calculated for all values of $d$.}
    \label{fig:synthetic_large_analysis}
\end{figure*}

In this sub-section we analyze the properties of the topologies included in the set of 900 synthetic networks (described in section \ref{sec:methods}) and contrast them against the properties of the real optical networks dataset.

\subsubsection{Structural Properties of Small Synthetic Network Dataset}
\label{sec:small_syn_network_connectivity_properties}

Initially, the diameters were calculated for all real optical networks and synthetic networks and plotted against the network size in Figure \ref{fig:synthetic_small_analysis}a. 

From Figure \ref{fig:synthetic_small_analysis}a one can see that most of real optical networks data (86\%) resides in the region $n\leq 40$. According to this Figure, the diameter grows with network size. This demonstrates that as the network size increases, networks become sparser, and their diameter increases. This observation is due to the grid-like structure of these networks. Figure \ref{fig:synthetic_small_analysis}a demonstrates that synthetic networks provide a broader range of diameters for testing when \( n \geq 40 \), compared to the singular samples for these network sizes. 

To further demonstrate this growth in sparsity in these networks, the network density was calculated and plotted in Figure \ref{fig:synthetic_small_analysis}b along with the number of edges in Figure \ref{fig:synthetic_small_analysis}d. Sparseness increases with network size, shown by the downward trending network density in Figure \ref{fig:synthetic_small_analysis}b. This results from the linear scaling in edges seen in Figure \ref{fig:synthetic_small_analysis}d. It can be seen how synthetic networks complement the real dataset, especially for larger networks.

Finally, the maximum edge betweenness - defined in \eqref{eq:edge_betweeness} - of the real and the synthetic networks was calculated and plotted in Figure \ref{fig:synthetic_small_analysis}c. Figure \ref{fig:synthetic_small_analysis}c shows maximum edge betweenness growing super-linearly with network size, indicating more paths crossing edges. In optical networking, this implies more wavelengths, higher costs, and allocation challenges. Therefore, as real optical networks are growing in size, it is important that researchers validate algorithmic solutions on a range of larger networks.

\subsubsection{Spatial Properties of Small Synthetic Network Dataset}
\label{sec:link_lengths}
The small synthetic dataset was generated with nodes scattered across a rectangle spanning most of the North American continent. To compare with real networks of similar scale, 16 networks spanning the North American continent were selected from the real dataset. Figure \ref{fig:usa_network_distances} shows the distribution of fibre lengths for these 16 real networks and the small synthetic networks.

The mean fibre lengths of the real North American networks (636 km) are slightly larger than those of the small synthetic dataset (593 km). The standard deviation is also larger for the real networks (480 km versus 415 km for synthetic networks). A KS2-test yielded a p-value of 0.0039, indicating that the two fibre length populations do not originate from the same distribution.
This difference in fibre lengths can be attributed to two factors. 
\begin{enumerate}
    \item \textbf{Node Placement:} In real optical networks, nodes are not necessarily uniformly placed throughout the 2D space, unlike in the synthetic dataset.
    \item \textbf{Network size influence:} As network size grows, average fibre lengths decrease for both real and synthetic networks. The North American dataset contains more smaller networks, skewing the distribution towards longer fibre lengths. In contrast, the synthetic dataset has an equal number of samples for each network size, resulting in a smaller mean and standard deviation for fibre lengths.
\end{enumerate}

\subsection{Large Synthetic Network Dataset (270,000 Networks)}

In this sub-section we analyze the 270,000 topologies included in the large synthetic dataset.

\subsubsection{Structural Properties of Large Synthetic Network Dataset}

To show the different network density profiles resulting from different values of $d$ used to generate the number of edges as $d \cdot n$, the network density and number of edges were calculated and plotted along with network size in Figure \ref{fig:synthetic_large_analysis}a and b, respectively. When $d$ = 1.2 the networks are very sparse and, therefore, the networks sometimes have more edge additions, as to ensure bi-connectivity network resilience constraints. These network density trends mirror those seen in real networks in Figure \ref{fig:synthetic_small_analysis}b.

The diameters of the synthetic networks were calculated for each value of $d$ and their probability distribution plotted in
Figure \ref{fig:synthetic_large_analysis}c. Empirically, some $d$ values provide similar diameter distributions, resulting in approximately 4 distinct sets across the 10 $d$ values. This dataset offers a diverse range of diameters between 1 and 30 for researchers to choose from.

\subsubsection{Spatial Properties of Large Synthetic Network Dataset}
The large synthetic networks dataset distributes nodes over three different distance scales to diversify the resultant fibre lengths. The fibre
length distributions of the three distance scales are calculated and plotted in Figure  \ref{fig:synthetic_large_analysis}d. From left to right, the distributions increase in average length and standard deviation, corresponding to small, medium, and large node scattering. This variety in distance distributions provides users with more options when selecting topologies for their use-case, which is crucial given the significant impact of fibre lengths on optical transmission. 

Given these real and synthetic network datasets, there is a remaining question of how does one choose which networks to use for benchmarking. The following section will investigate how to select networks from the real and synthetic network datasets for benchmarking.

\section{Topology Selection}
\label{sec:topology_selection}
Selecting from a large dataset of diverse topologies can be a difficult task, particularly for real topologies. In this section we discuss effective mechanisms for topology selection.
\subsection{Clustering of Real Networks}
Real-world topology selection is facilitated by applying K-means clustering to group real-world graphs based on their structural, spatial, and spectral properties. There are two optimisations in the process: (i) Maximisation of feature uniqueness using Principal Component Analysis, and (ii) Maximisation of decision boundary distance among K-means generated clusters using Support Vector Machines (SVM). The optimal metric selection is based on (i), and the number of clusters is selected based on (ii). Once optimal metrics are decided for a combined clustering approach, Principal Component Analysis (PCA) is used to reduce vector dimensionality, K-means is used to create the clusters and SVM classifiers generate decision boundaries. 
The resulting clusters  and topology assignments are documented in Appendix \ref{appendix:A}. This clustering approach represents an initial step towards a systematic selection of networks for benchmarking algorithmic solutions. The sklearn~\cite{Scikit-Learn} library is used for all tasks in the pipeline including dimensionality reduction, clustering, and decision boundary identification. 

\subsubsection{Levels of Clustering: Structural, Spectral, Spatial and Combined}
Clusters were computed at four different levels (structural, spatial, spectral, and combined). Each level captures different aspects of network characteristics: structural metrics reveal structural similarities, spectral metrics uncover similarities in the 
spectrum, spatial metrics highlight geographic and distance-related similarities of an optical network, and finally, combined metrics provide a holistic view, integrating all aspects.

\subsubsection{Principal Component Analysis and Metric Selection: Maximising Feature Uniqueness for Individual Clustering Level} 
The aim behind selecting particular features from each category was to optimise the final combined clustering, which is further used for topology characterisation and selection. Topology selection based only on one category is biased, and thus, we proposed this approach. We employed PCA for metric selection, focusing on maximising feature uniqueness across individual clustering levels. We outline the process for selecting a subset of metrics from a total of 21 graph metrics, the same set of features used in correlation analysis. 
\begin{itemize}[noitemsep]
    \item \textbf{Number of Metrics Per Level Set to 3:} We identified the smallest number of features to be chosen from each category. This was constrained by the number of spectral metrics under consideration which was three. To ensure equal and fair representation, we systematically searched for 3 metrics from each individual category to be aggregated in the combined cluster. This was applied across all domains, with 3 metrics each selected from the available structural (11 options), and spatial (7 options) metrics. 
    \item \textbf{Metric Selection Principle:} Our objective was to identify features that are least reducible by PCA \cite{daultrey1976principal}, implying they contain unique information. This unconventional use of PCA allows us to maximise feature uniqueness and minimise redundancy, potentially capturing a wider range of network characteristics. 
We systematically evaluated PCA for all possible combinations of three variables within spectral, structural, and spatial domains and the optimal metrics were selected based on the lowest explained variance in 2 dimensional PCA (0.86, 0.99, and 0.98 respectively). Only 1 combination was possible for spectral metrics. While 0.99 is still high, it was the lowest we found among all combinations of structural metrics. Furthermore, if there were groups of 3 variables with the same explained variance, we chose the lesser correlated features based on our previous analysis. 
\item \textbf{Maximise Feature Diversity:} PCA was applied to spectral, spatial, structural, and combined clustering levels. Explained variance ratios, quantifying preserved information, are presented in Table \ref{tab:clustering_metrics_summary}. 
The combined metrics (9 dimensions) retain 76\% variance in 2 dimensions, effectively condensing information (Table \ref{tab:pca_results}). Our methodology of selecting features with lower combined explained variance, ensures diverse clustering inputs, capturing a wide range of network characteristics for comprehensive analysis.
\end{itemize}

\begin{table}[ht]
\centering
\resizebox{\columnwidth}{!}{
\begin{tabular}{|c|c|c|c|}
\hline
\textbf{Metrics} & \textbf{PC1 (\%)} & \textbf{PC2 (\%)} & \textbf{Cumulative Variance (\%)} \\ \hline
Spectral   & 47\% & 36\% & 83\% \\ \hline
Spatial    & 83\% & 15\% & 98\% \\ \hline
Structural & 87\% & 12\% & 99\% \\ \hline
Combined   & 49\% & 26\% & 76\% \\ \hline
\end{tabular}
}
\caption{Cumulative variance explained by PC1 and PC2 for different metrics.}
\label{tab:pca_results}
\end{table}

\begin{table*}[!t]
\centering
\begin{tabular}{|c|c|}
\hline
\textbf{Type of Metric} & \textbf{Names} \\ \hline
\textbf{Structural} & Diameter (Hops), Average Shortest Path Length (Hops), Network Density  \\ \hline
\textbf{Spectral} & Normalized Spectral Radius, Normalised Algebraic Connectivity, Normalised Weighted Spectral Distribution \\ \hline
\textbf{Spatial} & Normalized Average Link Length, Normalized Diameter, Normalized Average Shortest Path Length (km) \\ \hline
\end{tabular}
\caption{
Optimal metrics are obtained at each Structural, Spatial, and Spectral categories by maximising feature diversity. 
}
\label{tab:metric_types}
\end{table*}

\subsubsection{K-means Clustering: Number of Clusters Selection and Evaluation By Maximising Margin for Combined Clustering}
\begin{figure*}[h!]
    \centering
    \includegraphics[width=\linewidth]{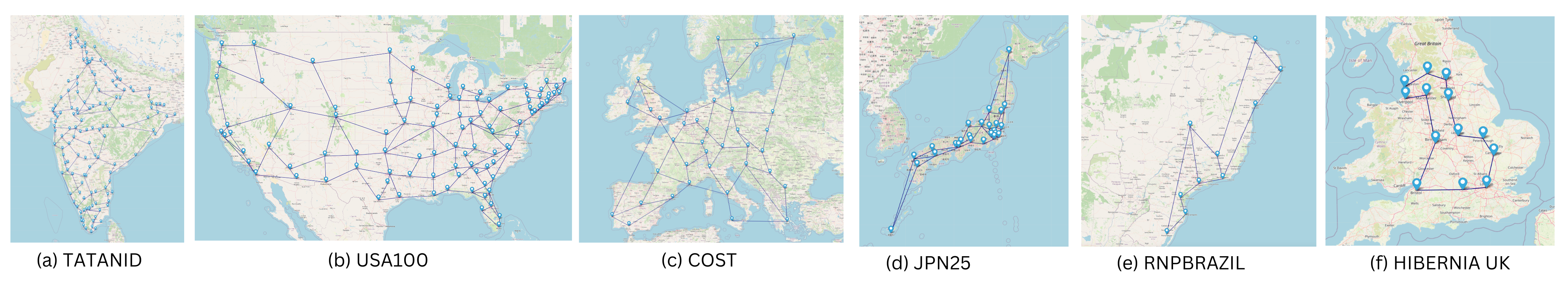}
    \caption{Topology visualisations of 2 networks from each combined-level cluster: (a) TATANID (India) and (b) USA100 (USA) in class 0, (c) COST (Europe) and (d) JPN25 (Japan) in class 1, (e) RNPBRAZIL (Brazil) and (f) HIBERNIA UK (UK) in class 2.}\label{fig:top_cluster_geo_visulaisation}
\end{figure*}

\begin{figure*}[h!]
    \centering
    \includegraphics[width=\linewidth]{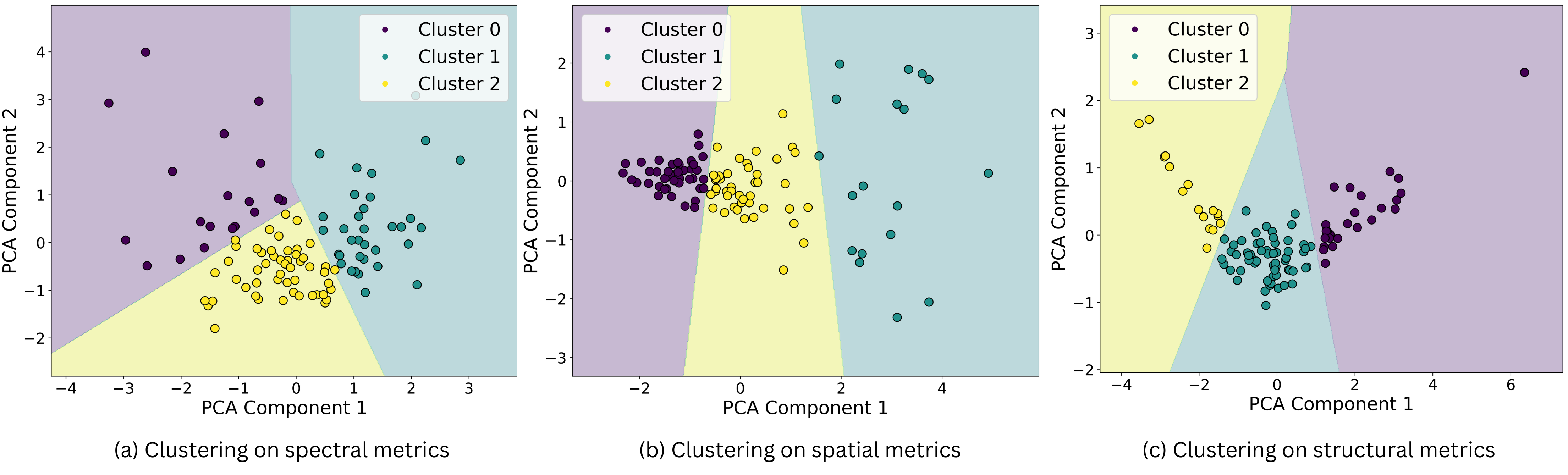}
    \caption{Clustering on the optimal features from each individual level—(a) spectral, (b) spatial, and (c) structural—helps in optical network selection. The diversity of clusters at each level strengthens the combined-level clustering.}
    \label{fig:clustering_individual}
\end{figure*}

\begin{figure}[h!]
    \centering
    \includegraphics[width=\linewidth]{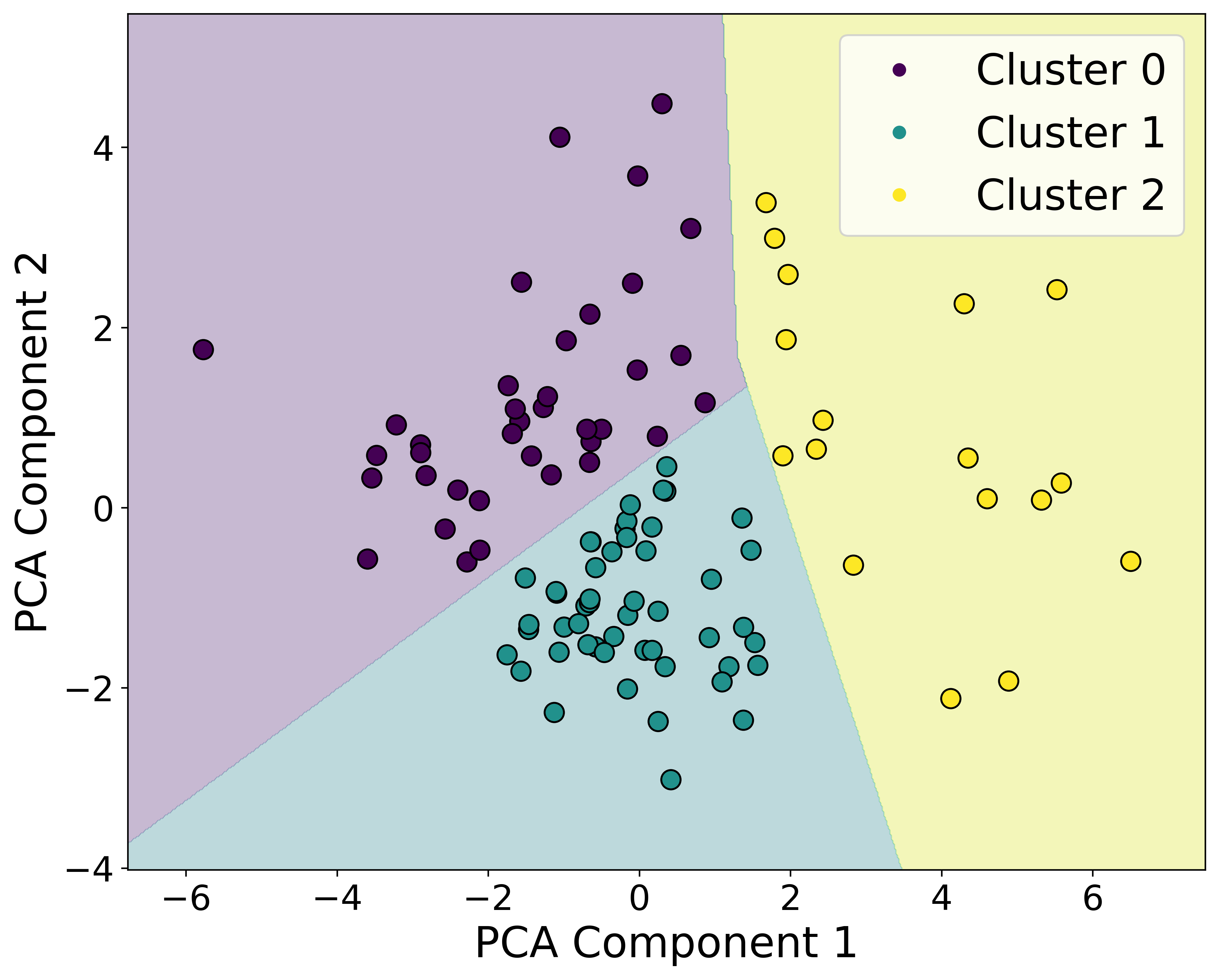}
    \caption{Clustering on combined metrics maximises the margin (0.27), achieves the greatest linear separability, and is recommended for diverse and objective topology selection.}
    \label{fig:clustering_combined_metrics}
\end{figure}

Given the absence of inherent classification schemes for network topologies, unsupervised learning presents a valuable approach for exploring and categorising the diverse landscape of optical networks.
\begin{itemize}[noitemsep]
    \item \textbf{Feature Selection and Preparation:} We selected three features from each level (structural, spatial, and spectral), resulting in nine representative features. 
To ensure comparability, all features were standardized.
    \item \textbf{Clustering Process:} We employ a three-step clustering process: (i) \textit{PCA} is used for dimensionality reduction to 2 dimensions. (ii) \textit{K-Means++ clustering} is applied \cite{hartigan1979k} in the reduced space for efficient centroid initialization and convergence. 
    (iii) The \textit{SVM} \cite{platt1999probabilistic} with a linear kernel aims to find a hyperplane that best separates the high-dimensional data points in the 2 dimensional PCA-transformed space. Clusters are treated as pseudo-labels, with the classifier evaluated on its ability to predict these labels. 
\item \textbf{Evaluation Criteria:} 
Clusters are evaluated using:
    \begin{itemize}[noitemsep]
    \item \textbf{Linear Separability:} We ensured clusters could be separated by linear functions, evaluated using SVM.
    \item \textbf{Margin Maximisation:} We optimised decision boundaries to maximise the distance between them, as determined by SVM.
\end{itemize}
    \item \textbf{Cluster Optimisation Process:}
    We evaluated configurations ranging from 3 to 7 clusters, determining the optimal number by maximizing the margin for the combined clustering level (Table \ref{table:clusters_and_margin}). Three clusters proved optimal for combined, spectral, and structural levels, while four clusters slightly outperformed for spatial metrics. For consistency and comparability, we applied three clusters across all individual clustering levels.
\item \textbf{Cluster Visualisation:} We visualised each cluster and their Support Vector Classification (SVC) boundaries for each cluster size and found 3 to be suitable as per visual and boundary analysis as well. In the next step, hyperplanes were calculated and plotted for spectral, spatial, and structural metrics (Figures \ref{fig:clustering_individual}a, b, c), as well as for the combined 9 metrics (Figure \ref{fig:clustering_combined_metrics}).
    \item \textbf{Clustering Evaluation:} If an SVM with a linear kernel can separate the clusters with high accuracy, it implies that the clusters are linearly separable. We computed the ratio of correctly separated samples by the Support Vector Classification (SVC) which is quantified by accuracy. While not all good clusterings are necessarily linearly separable, when they are, it can indicate well-defined and distinctly separated clusters. The combined clustering achieved 100\% accuracy with the highest margin of 0.277, as shown in Table \ref{tab:clustering_metrics_summary}. 
    This demonstrates that the combined cluster, leveraging strategically selected features, maximizes the margin compared to individual clusters.
    \item \textbf{Topology Selection with Combined Clustering Results:} The combined approach (Figure \ref{fig:clustering_combined_metrics}) reveals patterns that emerge only when considering multiple aspects simultaneously. It mitigates limitations and biases of individual metric types, showing clearer separation between clusters compared to individual metric-based clustering (Figure \ref{fig:clustering_individual}). The reduced overlap suggests that combining metrics from different domains (structural, spatial, spectral) provides greater discriminative power. Therefore, combined clustering approach demonstrates the highest quality in terms of linear separability, quantified through margin and is used to categorise networks for topology selection (Appendix \ref{appendix:A}). Figure \ref{fig:top_cluster_geo_visulaisation} visualises different topologies selected from each cluster identified by the combined clustering approach.
\end{itemize}

\begin{table*}[h!]
\centering
\begin{tabular}{|c|c|c|c|c|}
\hline
\textbf{Number of Clusters} & \textbf{Spectral Margin} & \textbf{Spatial Margin} & \textbf{Structural Margin} & \textbf{Combined Margin} \\ \hline
3 & 0.2212 & 0.2405 & 0.2378 & \textbf{0.2770} \\ \hline
4 & 0.1858 & 0.2621 & 0.2026 & 0.2431 \\ \hline
5 & 0.1520 & 0.2095 & 0.2110 & 0.1904 \\ \hline
6 & 0.1399 & 0.1811 & 0.1806 & 0.1883 \\ \hline
7 & 0.1223 & 0.1613 & 0.1465 & 0.1631 \\ \hline
\end{tabular}
\caption{Optimal number of clusters identified by maximising decision boundary margin at the combined level clustering.}
\label{table:clusters_and_margin}
\end{table*}

\begin{table}[ht]
\centering
\resizebox{\columnwidth}{!}{
\begin{tabular}{|c|c|c|c|c|}
\hline
\textbf{Metrics} & \textbf{Dimensions} & \textbf{Variance of PC1 + PC2} & \textbf{Accuracy} & \textbf{Margin} \\ \hline
Spectral & 3 & 0.83 & 0.981 & 0.221 \\ \hline
Spatial & 3 & 0.98 & 1.000 & 0.232 \\ \hline
Structural & 3 & 0.99 & 1.000 & 0.238 \\ \hline
Combined & 9 & 0.76 & 1.000 & \textbf{0.277} \\ \hline
\end{tabular}
}
\caption{Variance explained by first two principle components, accuracy and margin for 4 different levels of clustering.}
\label{tab:clustering_metrics_summary}
\end{table}

\subsubsection{Real Topology Selection}
\begin{figure}[t!]
    \centering
    \clearpage{}\begin{tikzpicture}

\definecolor{crimson2143940}{RGB}{214,39,40}
\definecolor{darkgray176}{RGB}{176,176,176}
\definecolor{darkorange25512714}{RGB}{255,127,14}
\definecolor{forestgreen4416044}{RGB}{44,160,44}
\definecolor{goldenrod18818934}{RGB}{188,189,34}
\definecolor{gray127}{RGB}{127,127,127}
\definecolor{lightgray204}{RGB}{204,204,204}
\definecolor{mediumpurple148103189}{RGB}{148,103,189}
\definecolor{orchid227119194}{RGB}{227,119,194}
\definecolor{sienna1408675}{RGB}{140,86,75}
\definecolor{steelblue31119180}{RGB}{31,119,180}

\begin{axis}[
legend cell align={left},
legend style={
  fill opacity=0.8,
  draw opacity=1,
  text opacity=1,
  at={(0.08,1.05)},
  anchor=south west,
  draw=lightgray204
},
tick align=outside,
tick pos=left,
x grid style={darkgray176},
xlabel={PCA 1},
xmajorgrids,
xmin=-0.903953392821643, xmax=1.01963834985251,
xtick style={color=black},
y grid style={darkgray176},
ylabel={PCA 2},
ymajorgrids,
ymin=-0.398145392721969, ymax=0.902861163414197,
ytick style={color=black}
]
\addplot [only marks, draw=steelblue31119180, fill=steelblue31119180, mark=*]
table{-0.788397 0.436793
};
\addlegendentry{Diameter (hops)}
\addplot [only marks, draw=darkorange25512714, fill=darkorange25512714, mark=*]
table{-0.816517 0.432017
};
\addlegendentry{Average Shortest Path Length (Hops)}
\addplot [only marks, draw=forestgreen4416044, fill=forestgreen4416044, mark=*]
table{0.931184 -0.118972
};
\addlegendentry{Edge Density}
\addplot [only marks, draw=crimson2143940, fill=crimson2143940, mark=*]
table{0.336954 0.321519
};
\addlegendentry{Spectral Radius}
\addplot [only marks, draw=mediumpurple148103189, fill=mediumpurple148103189, mark=*]
table{0.801492 -0.339009
};
\addlegendentry{Algebraic Connectivity}
\addplot [only marks, draw=sienna1408675, fill=sienna1408675, mark=*]
table{-0.026869 0.843725

};
\addlegendentry{WSD}
\addplot [only marks, draw=orchid227119194, fill=orchid227119194, mark=*]
table{0.932202 0.230982
};
\addlegendentry{Average Link Length (km)}
\addplot [only marks, draw=gray127, fill=gray127, mark=*]
table{0.451127 0.771938
};
\addlegendentry{Diameter (km)}
\addplot [only marks, draw=goldenrod18818934, fill=goldenrod18818934, mark=*]
table{0.683316 0.629842
};
\addlegendentry{Average Shortest Path Length (km)}
\end{axis}

\end{tikzpicture}
\clearpage{}
    \caption{Correlation between PCA 1 and 2 components and the graph metrics used for clustering.}
    \label{fig:pca_corr}
\end{figure}
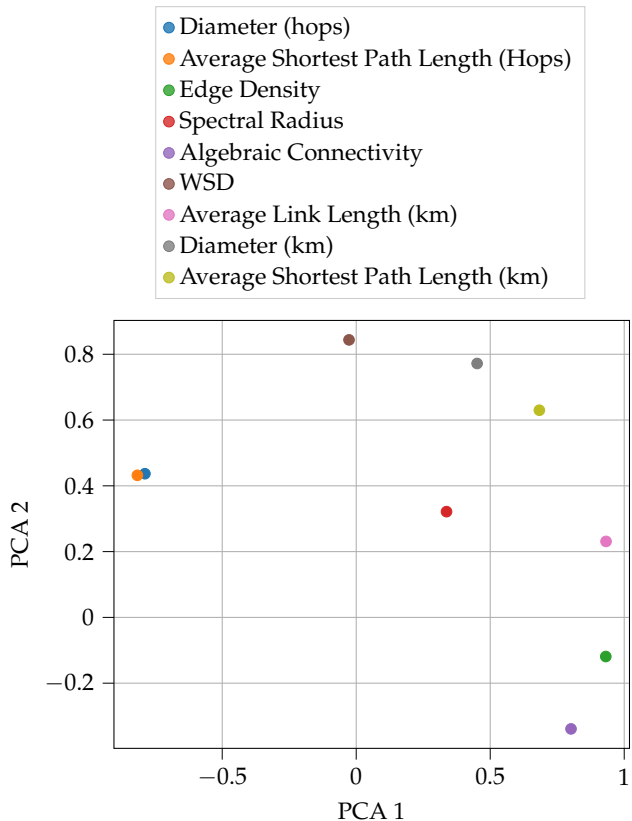

Given the combined clustering in Figure \ref{fig:clustering_combined_metrics}, it is difficult to understand the abstract quantities of the components of the PCA (x/y-axis). To make the clusters and the location of the networks more interpretable, a linear correlation analysis between the PCA components and the graph metrics are performed using the Pearson's correlation coefficient. The correlation is calculated between each component and metric and plotted in Figure \ref{fig:pca_corr}. For this analysis, correlations higher than 0.7 are considered.

In Figure \ref{fig:pca_corr}, one can see that PCA 1 clearly correlates well to the algebraic connectivity, normalised average link length and network density with correlation coefficients of 0.80, 0.93 and 0.93 respectively. This means that metrics on the right hand side of Figure \ref{fig:clustering_combined_metrics} generally will be better connected with larger average link lengths. PCA 2 has a very clear correlation between WSD and normalised diameters with correlation coefficients of 0.84 and 0.77 respectively. This means that networks higher up on the y-axis of Figure \ref{fig:clustering_combined_metrics} will tend to have larger WSD (more sparse) and larger spatial diameters. 

Taking these correlations into account we can make some recommendations for a network selection process that ensures diversity in network size, structural distributions and spatial distributions. To do this at least two samples from each cluster should be selected. This is a minimum choice to capture metric diversity from within clusters. PCA 1 is negatively correlated ($\rho=-0.77$) with network size, therefore all chosen samples should be spaced out along the x-axis of Figure \ref{fig:clustering_combined_metrics}. This will allow for sampling of diverse network density values. To ensure structural and spatial diversity, these samples should also be spread out across the y-axis capturing different values of WSD and spatial diameters. Selecting the networks evenly across this space and from each cluster will ensure that the networks reflect diversity in terms of network size, structural and spatial properties. 

The following section briefly looks at the selection problem for mainly the large synthetic network dataset.

\subsection{Selecting Synthetic Networks}

For the small synthetic network dataset, the selection process is straight forward. Since the networks have been sampled and generated from a single distribution, it is simply a matter of the user determining the range of network sizes that they would like to test and the number of samples per network size (up to 10). However, for the large synthetic network dataset there are more choices to be made. In addition to network size, there are different distance and network density scales to choose from. This can be navigated by choosing a range of network sizes, a singular or multiple values of $d$ and the distance scale which one wishes to use. However, due to the range of topologies produced one might want to select subsets of distributions from network size, spatial and structural properties, depending on requirements.

\begin{figure}[t!]
    \centering
    \includegraphics[scale=0.8]{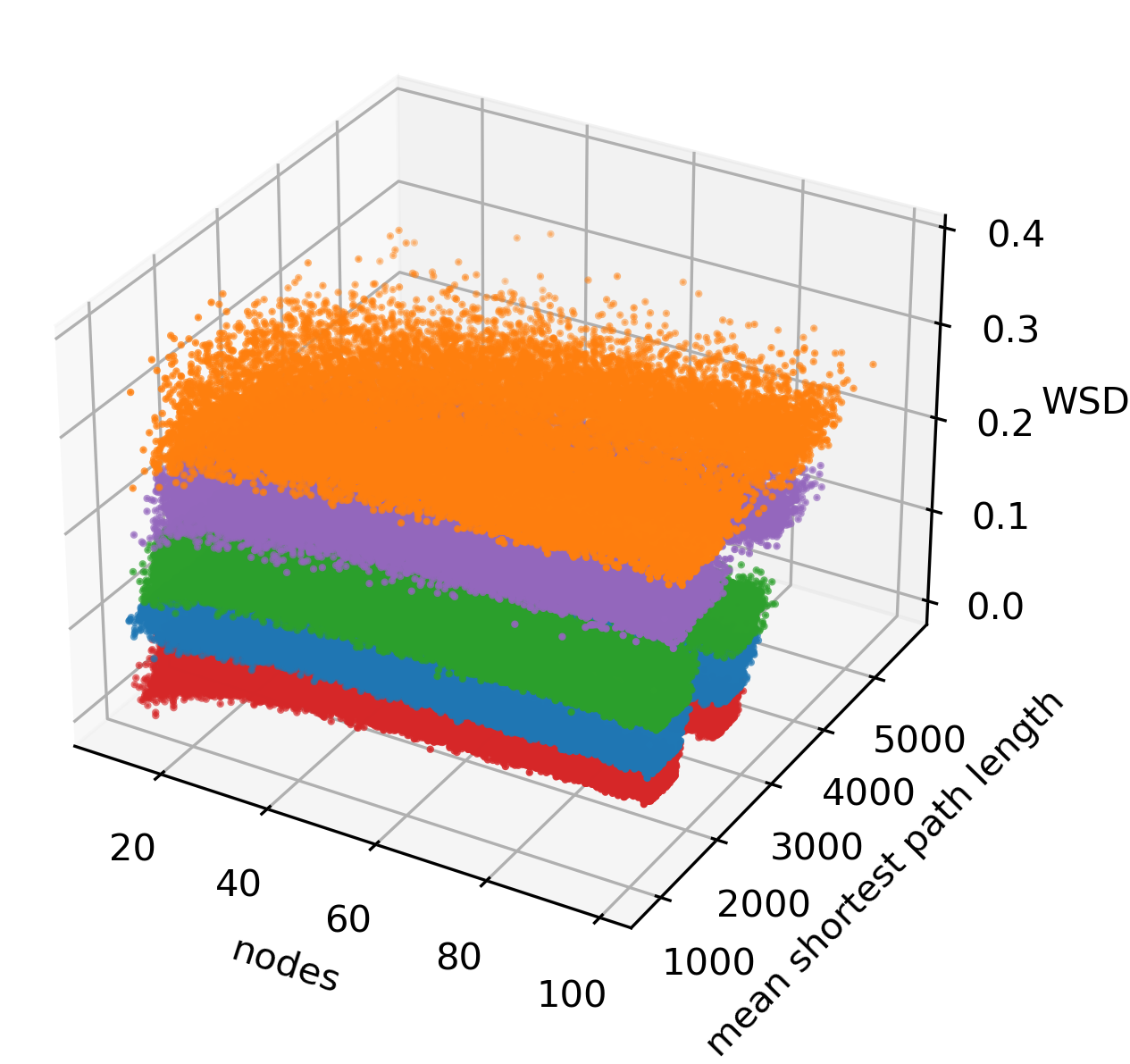}
    \caption{WSD clustering for large synthetic network dataset plotted along with network size and mean shortest path length.}
    \label{fig:cluster_syn_large}
\end{figure}

To achieve this, for each sample within the large synthetic dataset the network size, mean shortest path length (spatial property) and WSD (structural property) is calculated and plotted in Figure \ref{fig:cluster_syn_large}. The different colors represent different k-means clusters of the data, according to WSD. This is done to demonstrate different slices of networks. The WSD-axis in Figure \ref{fig:cluster_syn_large} represents the diversity in terms of structure. Here smaller values represent networks that are better connected, either with larger network densities, or via better connected structures (smaller diameter/path lengths). 

Using these three metrics, the user can determine appropriate values for each of these and slice the data accordingly to their needs. An example ML use-case might be as follows. A user has trained a model to perform a resource allocation task. They wish now to see how their model generalises to other distributions. Therefore, they sample 900 networks from the large synthetic dataset, encompassing 18 network sizes, 10 WSD values and 5 mean shortest path lengths, chosen uniformly across each distribution.  The same process could be taken to sample networks for training a ML model.

\section{Conclusion} 
We introduced Topology Bench, a comprehensive dataset of real and synthetic optical network topologies. This paper focused on three main outcomes: (a) standardised real and synthetic network datasets (b) clustering of datasets for selection (c) network analysis  quantifying resource use, throughput and resilience. These topologies will be the starting point for a variety of combinatorial optimisation problems such as physical topology design and resource and wavelength allocation and a multitude of network modelling problems for the future. 
Our study presents a uniform, spatially-referenced, geographically diverse dataset of 105 real-world optical network topologies, significantly expanding upon existing resources by 61.5\%. The dataset integrates and enhances topologies from established repositories (SNDlib, Internet Topology Zoo, Topohub) and incorporates networks often referenced (but not openly published) in academic literature. A key contribution is the manual geocoding of all networks to compute fibre link lengths, ensuring comprehensive spatial coordinate data—a feature often absent or incomplete in prior datasets.
\newline\newline
To complement the set of real topologies, and to address the diverse requirements of researchers studying various optimisation problems in optical networking, we employed the SNR-BA generative model to create two synthetic datasets, differentiated on size. The SNR-BA model was first validated using node locations from real networks. We then generated a small synthetic dataset of 900 networks to complement the real network set, providing a manageable corpus for initial studies and algorithm development. Recognising the growing demand for large-scale data in machine learning applications, we subsequently created an extensive synthetic dataset comprising 270,000 networks. This larger dataset offers a wide range of network sizes, spatial properties, and connectivities, enabling comprehensive analysis and robust training of machine learning models for optical network optimisation.
\newline\newline
Finally, to mitigate the inherent subjectivity in topology selection for network modelling—often influenced by researcher bias, data accessibility constraints, or assumptions of appropriateness—we applied unsupervised machine learning on Topology Bench. 
K-means clustering was applied to the set of 105 real optical networks using 9 diverse metrics, resulting in 3 distinct groups. These clusters were visualized through principal component analysis, and correlations between the reduced dimensions and the metrics were examined. The analysis revealed significant correlations between network properties: network density, average fibre length, and algebraic connectivity aligned with one principal component, while weighted spectral distribution and diameter corresponded to the other.
Based on this, users can choose networks that are diverse in terms of structure (network density), spectrum (weighted spectral distribution, algebraic connectivity) and spatial properties (average fibre lengths, diameter). 
The approach to topology selection based on structural, spatial and spectral properties was also introduced for the choice of  distributions of networks from the large synthetic network dataset. Topology Bench provides optical core network researchers with a systematic graph theoretic approach to select a set of real topologies and synthetic networks for benchmarking and future machine learning applications to optical network design.

\section*{Acknowledgments}
Financial support from UK EPSRC Doctoral Training Programme, a UK EPSRC / BT iCASE studentship (EP/W522120/1), UK EPSRC grant  the Centre for Doctoral Training in Connected Electronic and Photonic Systems (EP/S022139/1) and the UK EPSRC Programme Grant TRANSNET (EP/R035342/1)  as well as the  Microsoft 'Optics for the Cloud' programme, is gratefully acknowledged.

\section*{Data access}
The dataset for Topology Bench is hosted at \href{https://zenodo.org/records/13921775}{Zenodo} and the code for analysis is available on \href{https://github.com/TopologyBench}{GitHub}. For the purpose of open access, the author has applied a Creative Commons Attribution (CC BY) license to any Author Accepted Manuscript version arising. We believe it would be useful to expand the data set to include metro, access and data centre networks although the inclusion of geolocation data may require manual intervention. 
To contribute your dataset or topology to our database, please contact the authors.

\bibliography{JOCN-Template}

\newpage
\onecolumn
\appendix
\section{Appendix: Topology Bench Real Network Topologies}
\label{appendix:A}
\nopagebreak

\centering
\scriptsize
\setlength{\tabcolsep}{3pt}
\begin{longtable}[!htbp]{|>{\centering\arraybackslash}p{0.02\textwidth}|>{\centering\arraybackslash}p{0.15\textwidth}|>{\centering\arraybackslash}p{0.02\textwidth}|>{\centering\arraybackslash}p{0.02\textwidth}|>{\centering\arraybackslash}p{0.07\textwidth}|>{\centering\arraybackslash}p{0.07\textwidth}|>{\centering\arraybackslash}p{0.09\textwidth}|>{\centering\arraybackslash}p{0.09\textwidth}|>{\centering\arraybackslash}p{0.10\textwidth}|>{\centering\arraybackslash}p{0.07\textwidth}|>{\centering\arraybackslash}p{0.07\textwidth}|>{\centering\arraybackslash}p{0.05\textwidth}|}

\caption{Core optical network topologies real data set in Topology Bench. Number of nodes and number of edges are denoted by $n$ and $m$ respectively.}\label{tab:topo_data_part1} \\
\hline
\thead{No.} & \thead{Topology\\Name} & \thead{$n$} & \thead{$m$} & \thead{Avg Node\\Degree} & \thead{Diameter\\(Hops)} & \thead{Avg Link\\Length (km)} & \thead{Algebraic\\Connectivity} & \thead{Region} & \thead{Cluster\\Label} & \thead{Survivable} & \thead{Outlier}\\
\hline
\endfirsthead

\hline
\thead{No.} & \thead{Topology\\Name} & \thead{$n$} & \thead{$m$} & \thead{Avg Node\\Degree} & \thead{Diameter\\(Hops)} & \thead{Avg Link\\Length (km)} & \thead{Algebraic\\Connectivity} & \thead{Region} & \thead{Cluster\\Label} & \thead{Survivable} & \thead{Outlier}\\
\hline
\endhead

\hline
\multicolumn{11}{r}{Continued on next page} \\
\endfoot

\hline
\endlastfoot

   1 &             ABILENE &     11 &     14 &             2.55 &                5 &               1353.28 &                0.13 & USA & Class 1 & \cmark & \xmark\\
   2 &                 ANS &     18 &     25 &             2.78 &                6 &               1358.61 &                0.07 & USA & Class 2 & \xmark & \xmark\\
   3 &               ARNES &     17 &     20 &             2.35 &                6 &                 51.55 &                0.10 & Slovenia & Class 2 & \cmark & \xmark\\
   4 &             ARPANET &     20 &     32 &             3.20 &                6 &               1173.06 &                0.10 & USA & Class 2 & \cmark & \xmark\\
   5 &              ATMNET &     21 &     22 &             2.10 &                9 &                826.88 &                0.05 & USA & Class 0 & \xmark& \xmark\\
   6 &           BBNPLANET &     27 &     28 &             2.07 &                7 &                853.54 &                0.03 & USA & Class 2 & \xmark& \xmark\\
   7 &    BEYONDTHENETWORK &     29 &     41 &             2.83 &                6 &                888.67 &                0.05 & USA & Class 2 & \xmark& \xmark\\
   8 &                BICS &     33 &     48 &             2.91 &                8 &                610.97 &                0.04 & Europe & Class 2 & \xmark& \xmark\\
   9 &              BIZNET &     28 &     32 &             2.29 &               15 &                155.34 &                0.01 & Indonesia & Class 0 & \xmark& \xmark\\
  10 &                BREN &     10 &     11 &             2.20 &                4 &                170.27 &                0.17 & Bulgaria & Class 1 & \cmark & \xmark\\
  11 &             CANARIE19 &     19 &     26 &             2.74 &                7 &                923.11 &                0.08 & Canada/USA & Class 2 & \cmark & \xmark\\
  12 &             CANARIE24 &     24 &    33 &             2.75 &                7 &                966.22 &                0.051 & Canada/USA & Class 2 & \xmark& \xmark\\
  13 &              CERNET &     36 &     51 &             2.83 &                5 &                855.46 &                0.08 & China & Class 2 & \xmark& \xmark\\
  14 &              CESNET &     12 &     19 &             3.17 &                4 &                136.97 &                0.20 & China & Class 2 & \cmark & \cmark \\
  15 &            CLARANET &     15 &     18 &             2.40 &                4 &                659.69 &                0.25 & Europe & Class 2 & \xmark& \xmark\\
  16 &           CONUS100 &    100 &    136 &             2.72 &               15 &               1380.74 &                0.02 & USA & Class 0 & \cmark & \cmark \\
  17 &            CONUS30 &     30 &     36 &             2.40 &               11 &                780.28 &                0.03 & USA & Class 0 & \cmark & \xmark\\
  18 &           CONUS6077 &     60 &     77 &             2.57 &               14 &                521.00 &                0.02 & USA & Class 0 & \cmark & \xmark\\
  19 &           CONUS6079 &     60 &     79 &             2.63 &               15 &                543.87 &                0.02 & USA & Class 0 & \cmark & \xmark\\
  20 &             CONUS75 &     75 &     99 &             2.64 &               17 &                494.35 &                0.01 & USA & Class 0 & \cmark & \xmark\\
  21 &             CORONET &     60 &     79 &             2.63 &               15 &                543.87 &                0.02 & USA & Class 0 & \cmark & \xmark\\
  22 &                COST &     37 &     57 &             3.08 &                8 &                648.17 &                0.04 & Europe & Class 2 & \cmark & \xmark\\
  23 & CRLNETWORKSERVICES &     33 &     38 &             2.30 &               10 &                924.96 &                0.02 & USA & Class 0 & \xmark& \xmark\\
  24 &                CWIX &     23 &     29 &             2.52 &                7 &                922.32 &                0.07 & USA & Class 2 & \xmark& \xmark\\
  25 &          DARKSTRAND &     28 &     31 &             2.21 &               11 &                681.40 &                0.04 & USA & Class 0 & \cmark & \xmark\\
  26 &               DIGEX &     31 &     35 &             2.26 &               10 &                803.30 &                0.04 & USA & Class 0 & \cmark & \xmark\\
  27 &                DT17 &     14 &     22 &             3.14 &                5 &                185.58 &                0.20 & Germany & Class 2 & \cmark & \xmark\\
  28 &         ELIBACKBONE &     20 &     30 &             3.00 &                6 &               1056.60 &                0.10 & USA & Class 2 & \cmark & \xmark\\
  29 &                EON &     19 &     37 &             3.89 &                5 &                918.67 &                0.14 & Europe & Class 2 & \cmark & \cmark \\
  30 &              EPOCH &      6 &      7 &             2.33 &                3 &               2277.42 &                0.27 & USA & Class 1 & \cmark & \cmark \\
  31 &              ERNET &     16 &     18 &             2.25 &                4 &                897.72 &                0.19 & India & Class 2 & \xmark& \xmark\\
  32 &              FUNET &     24 &     27 &             2.25 &                9 &                157.81 &                0.06 & Finland & Class 0 & \xmark& \xmark\\
  33 &             GAMBIA &     12 &     12 &             2.00 &                5 &                 61.57 &                0.10 & Gambia & Class 1 & \xmark& \xmark\\
  34 &             GBLNET &      8 &      7 &             1.75 &                4 &                818.59 &                0.27 & Europe & Class 1 & \xmark& \cmark \\
  35 &              GEANT &     22 &     36 &             3.27 &                5 &               1388.75 &                0.15 & Inter-Continental& Class 2 & \cmark & \xmark\\
  36 &             GEANT2 &     35 &     55 &             3.14 &                8 &                881.15 &                0.07 & Europe & Class 2 & \cmark & \xmark\\
  37 &          GERMANY50 &     50 &     88 &             3.52 &                9 &                193.14 &                0.05 & Germany & Class 2 & \cmark & \xmark\\
  38 &             GETNET &      7 &      8 &             2.29 &                3 &               2146.64 &                0.57 & USA & Class 1 & \xmark& \cmark \\
  39 &               GRNET &     34 &     39 &             2.29 &                8 &                190.21 &                0.14 & Greece & Class 2 & \xmark& \xmark\\
  40 &  GTSCZECHREPUBLIC &     26 &     25 &             1.92 &               17 &                 92.29 &                0.01 & Czech Republic & Class 1 & \xmark& \cmark \\
  41 &           GTSPOLAND &     26 &     28 &             2.15 &               12 &                154.73 &                0.05 & Poland & Class 0 & \xmark& \xmark\\
  42 &      HIBERNIACANADA &     10 &     10 &             2.00 &                7 &               1615.95 &                0.07 & Canada & Class 1 & \xmark& \xmark\\
  43 &      HIBERNIAGLOBAL &     53 &     76 &             2.87 &               18 &                719.31 &                0.01 & Global & Class 0 & \xmark& \xmark\\
  44 &    HIBERNIAIRELAND &      6 &      6 &             2.00 &                3 &                142.22 &                0.43 & Ireland & Class 0 & \xmark& \cmark \\
  45 &  HIBERNIANIRELAND &     15 &     16 &             2.13 &               10 &                 64.05 &                0.02 & Ireland & Class 1 & \xmark& \xmark\\
  46 &        HIBERNIAUK &     13 &     13 &             2.00 &                6 &                104.95 &                0.08 & UK & Class 1 & \cmark & \xmark\\
  47 & HIBERNIAUS & 20 & 27 & 2.70 & 10 & 405.30 & 0.01 & USA & Class 0 & \xmark& \xmark\\
48 & HOSTWAYINTERNATIONAL & 16 & 21 & 2.63 & 8 & 4159.84 & 0.05 & Global& Class 2 & \xmark& \cmark \\
49 & IBM & 18 & 24 & 2.67 & 6 & 1258.93 & 0.07 & USA & Class 2 & \xmark& \xmark\\
50 & INTEGRA & 27 & 36 & 2.67 & 7 & 757.45 & 0.09 & USA & Class 2 & \xmark& \xmark\\
51 & IRIS & 50 & 64 & 2.56 & 10 & 94.34 & 0.03 & USA & Class 0 & \xmark& \xmark\\
52 & ISTAR & 19 & 19 & 2.00 & 7 & 696.54 & 0.07 & Canada & Class 0 & \xmark & \xmark\\
53 & ITALY & 21 & 35 & 3.33 & 7 & 271.16 & 0.07 & Italy & Class 2 & \cmark & \xmark\\
54 & JANOSUS & 26 & 42 & 3.23 & 8 & 882.12 & 0.08 & USA & Class 2 & \cmark & \xmark\\
55 & JGN2PLUS & 11 & 10 & 1.82 & 7 & 598.39 & 0.03 & Japan & Class 1 & \xmark & \xmark\\
56 & JPN12 & 12 & 17 & 2.83 & 5 & 623.44 & 0.09 & Japan & Class 2 & \cmark & \xmark\\
57 & JPN25 & 25 & 43 & 3.44 & 8 & 318.70 & 0.03 & Japan & Class 2 & \cmark & \xmark\\
58 & JPN48 & 48 & 82 & 3.42 & 14 & 230.69 & 0.01 & Japan & Class 0 & \cmark & \xmark\\
59 & KAREN & 22 & 25 & 2.27 & 6 & 145.81 & 0.06 & NewZeland & Class 2 & \xmark& \xmark\\
60 & LAMBDARAIL & 19 & 23 & 2.42 & 7 & 907.53 & 0.07 & USA & Class 0 & \cmark & \xmark\\
61 & LAYER42 & 6 & 7 & 2.33 & 3 & 2056.05 & 0.73 & USA & Class 1 & \xmark& \cmark \\
62 & LONI & 32 & 36 & 2.25 & 12 & 76.72 & 0.03 & USA (Louisiana)& Class 0 & \cmark & \xmark\\
63 & MARWAN & 6 & 6 & 2.00 & 3 & 413.97 & 0.30 & Morocco & Class 1 & \cmark & \cmark \\
64 & MEMOREX & 19 & 24 & 2.53 & 7 & 205.93 & 0.06 & Europe & Class 2 & \cmark & \xmark\\
65 & METRONA & 33 & 41 & 2.48 & 11 & 640.21 & 0.01 & UK & Class 0 & \cmark & \xmark\\
66 & MZIMA & 14 & 18 & 2.57 & 5 & 1239.80 & 0.11 & USA & Class 2 & \cmark & \xmark\\
67 & NETRAIL & 7 & 10 & 2.86 & 2 & 1768.60 & 0.36 & USA & Class 1 & \cmark & \cmark \\
68 & NETWORKUSA & 35 & 39 & 2.23 & 10 & 130.86 & 0.02 & USA & Class 0 & \xmark& \xmark\\
69 & NEWNET & 26 & 31 & 2.38 & 9 & 787.70 & 0.05 & USA & Class 0 & \cmark & \xmark\\
70 & NEXTGEN & 16 & 16 & 2.00 & 13 & 661.50 & 0.01 & Australia & Class 0 & \xmark& \cmark \\
71 & NOBELEU & 28 & 41 & 2.93 & 8 & 622.18 & 0.05 & Europe & Class 2 & \cmark & \xmark\\
72 & NOBELGERMANY & 17 & 26 & 3.06 & 6 & 215.00 & 0.17 & Germany & Class 2 & \cmark & \xmark\\
73 & NOBELUS & 14 & 21 & 3.00 & 3 & 1464.78 & 0.27 & USA & Class 2 & \cmark & \xmark\\
74 & NOEL & 19 & 25 & 2.63 & 6 & 157.52 & 0.17 & USA & Class 2 & \xmark& \xmark\\
75 & NSFNET13 & 13 & 15 & 2.31 & 5 & 1480.27 & 0.22 & USA & Class 2 & \xmark& \xmark\\
76 & OPTUNETSWEDEN & 25 & 29 & 2.32 & 12 & 253.13 & 0.03 & Sweden & Class 0 & \xmark& \xmark\\
77 & OXFORD & 19 & 25 & 2.63 & 7 & 81.52 & 0.06 & USA & Class 2 & \cmark & \xmark\\
78 & PACKETEXCHANGE & 21 & 27 & 2.57 & 9 & 3410.12 & 0.02 & Global & Class 0 & \xmark& \cmark \\
79 & PALMETTO & 45 & 64 & 2.84 & 12 & 100.53 & 0.03 & USA & Class 0 & \xmark& \xmark\\
80 & PIONIER & 21 & 25 & 2.38 & 7 & 183.21 & 0.07 & Poland & Class 2 & \cmark & \xmark\\
81 & PIONIERL3 & 27 & 32 & 2.37 & 10 & 167.05 & 0.04 & Poland & Class 0 & \xmark& \xmark\\
82 & POLSKA & 12 & 18 & 3.00 & 4 & 282.11 & 0.21 & Poland & Class 2 & \cmark & \xmark\\
83 & PORTUGAL & 26 & 36 & 2.77 & 8 & 263.17 & 0.04 & Portugal & Class 2 & \cmark & \xmark\\
84 & PSINET & 24 & 25 & 2.08 & 11 & 769.97 & 0.02 & USA & Class 0 & \xmark& \xmark\\
85 & RAILTELINDIA & 19 & 28 & 2.95 & 6 & 649.37 & 0.18 & India & Class 2 & \cmark & \xmark\\
86 & REDCLARA & 14 & 17 & 2.43 & 4 & 2717.70 & 0.25 & South America & Class 2 & \xmark& \xmark\\
87 & REDIRIS & 18 & 30 & 3.33 & 4 & 506.26 & 0.20 & Spain & Class 2 & \cmark & \xmark\\
88 & RENATER & 27 & 35 & 2.59 & 6 & 231.56 & 0.12 & France & Class 2 & \cmark & \xmark\\
89 & RENATER1999 & 24 & 23 & 1.92 & 7 & 289.84 & 0.08 & France & Class 2 & \xmark& \cmark \\
90 & RENATER2001 & 24 & 27 & 2.25 & 6 & 273.86 & 0.13 & France & Class 2 & \xmark& \xmark\\
91 & RENATER2004 & 24 & 29 & 2.42 & 7 & 269.53 & 0.13 & France & Class 2 & \xmark& \xmark\\
92 & RENATER2006 & 26 & 34 & 2.62 & 7 & 247.94 & 0.10 & France & Class 2 & \xmark& \xmark\\
93 & RENATER2008 & 26 & 34 & 2.62 & 7 & 247.94 & 0.10 & France & Class 2 & \xmark& \xmark\\
94 & RENATER2010 & 37 & 48 & 2.59 & 9 & 215.25 & 0.07 & France & Class 2 & \xmark& \xmark\\
95 & RNPBRAZIL & 10 & 12 & 2.40 & 5 & 1049.74 & 0.18 & Brazil & Class 1 & \cmark & \xmark\\
96 & SAGO & 18 & 17 & 1.89 & 14 & 108.52 & 0.02 & USA & Class 0 & \xmark& \cmark \\
97 & SANET & 46 & 55 & 2.39 & 13 & 32.53 & 0.02 & Slovakia & Class 0 & \xmark& \xmark\\
98 & SANREN & 7 & 7 & 2.00 & 3 & 692.31 & 0.15 & South Africa & Class 1 & \cmark & \xmark\\
99 & SAVVIS & 19 & 20 & 2.11 & 8 & 811.75 & 0.04 & USA & Class 0 & \xmark& \xmark\\
100 & SPIRALIGHT & 15 & 16 & 2.13 & 8 & 132.21 & 0.07 & USA & Class 2 & \cmark & \xmark\\
101 & TATANID & 142 & 180 & 2.54 & 28 & 200.76 & 0.01 & India & Class 0 & \xmark& \cmark \\
102 & TELECOMSERBIA & 6 & 6 & 2.00 & 3 & 213.98 & 0.29 & Serbia & Class 1 & \cmark & \cmark \\
103 & UNIC & 15 & 17 & 2.27 & 8 & 72.73 & 0.04 & Denmark & Class 0 & \xmark& \xmark\\
104 & USA100 & 100 & 171 & 3.42 & 16 & 411.74 & 0.02 & USA & Class 0 & \cmark & \cmark \\
105 & VIA & 9 & 12 & 2.67 & 4 & 821.66 & 0.27 & Europe & Class 1 & \cmark & \xmark\\

\end{longtable}

\end{document}